\documentclass{article}
\usepackage{arxiv}
\usepackage{cite}
\usepackage{amsmath,amssymb,amsfonts}
\usepackage{algorithmic}
\usepackage{graphicx}
\graphicspath{{Images\_QPUF/}}
\usepackage{subcaption}
\usepackage{wrapfig}
\usepackage{textcomp}
\usepackage{xcolor}
\usepackage{lipsum} 
\usepackage{cite}
\usepackage{multirow}
\usepackage{url}
\usepackage{amsmath}
\usepackage{graphicx}
\usepackage[compact]{titlesec}
\usepackage{xcolor}
\usepackage{braket}
\usepackage{subcaption}
\usepackage{rotfloat}
\usepackage[T1]{fontenc}    
\usepackage{hyperref}       
\usepackage{url}            
\usepackage{amsmath}
\usepackage{amssymb}
\usepackage{algorithm}
\usepackage{algorithmic}
\usepackage{gensymb}
\usepackage{textcomp}
\usepackage{amsfonts}
\usepackage{booktabs}
\usepackage{siunitx}
\usepackage{caption}
\usepackage{amsmath}
\usepackage{hyperref}
\usepackage{subcaption}
\usepackage{balance}

\newcommand{\orcid}[1]{\href{https://orcid.org/#1}{\textcolor[HTML]{A6CE39}{\aiOrcid}}}
\usepackage{array}
\newcolumntype{P}[1]{>{\centering\arraybackslash}p{#1}}
\newcolumntype{M}[1]{>{\centering\arraybackslash}m{#1}}

\def\BibTeX{{\rm B\kern-.05em{\sc i\kern-.025em b}\kern-.08em
		T\kern-.1667em\lower.7ex\hbox{E}\kern-.125emX}}

\title{QPUF 2.0: Exploring Quantum Physical Unclonable Functions for Security-by-Design of Energy Cyber-Physical Systems
}

\author{ 
	\href{https://orcid.org/0000-0002-4487-3239}{\includegraphics[scale=0.06]{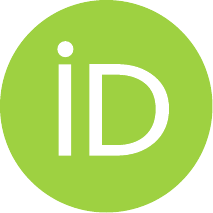}\hspace{1mm}Venkata K. V. V. Bathalapalli}  \\
	Department of Computer Science and Engineering\\
	University of North Texas\\
	\texttt{vb0194@unt.edu} \\
	\And
	\href{https://orcid.org/0000-0003-2959-6541}{\includegraphics[scale=0.06]{orcid.pdf}\hspace{1mm}Saraju P. Mohanty} \\
	Department of Computer Science and Engineering\\
	University of North Texas\\
	\texttt{saraju.mohanty@unt.edu} \\
	\And
	\href{https://orcid.org/0000-0001-9161-1728}{\includegraphics[scale=0.06]{orcid.pdf}\hspace{1mm}Chenyun Pan} \\
	Department of Electrical Engineering\\
	University of Texas at Arlington\\
	\texttt{chenyun.pan@uta.edu} \\
	\And
	\href{https://orcid.org/0000-0002-1616-7628}{\includegraphics[scale=0.06]{orcid.pdf}\hspace{1mm}Elias Kougianos} \\
	Department of Electrical Engineering\\
	University of North Texas\\
	\texttt{elias.kougianos@unt.edu}\\
}

\begin{document}

	\makeatletter
	\maketitle

	\begin{abstract}	 
		Sustainable advancement is being made to improve the efficiency of the generation, transmission, and distribution of renewable energy resources, as well as managing them to ensure the reliable operation of the smart grid. Supervisory control and data acquisition (SCADA) enables sustainable management of grid communication flow through its real-time data sensing, processing, and actuation capabilities at various levels in the energy distribution framework. The security vulnerabilities associated with the SCADA-enabled grid infrastructure and management could jeopardize the smart grid operations. This work explores the potential of Quantum Physical Unclonable Functions (QPUF) for the security, privacy, and reliability of the smart grid's energy transmission and distribution framework.   
		Quantum computing has emerged as a formidable security solution for high-performance computing applications through its probabilistic nature of information processing. This work has a quantum hardware-assisted security mechanism based on intrinsic properties of quantum hardware driven by quantum mechanics to provide tamper-proof security for quantum computing driven smart grid infrastructure. This work introduces a novel QPUF architecture using quantum logic gates based on quantum decoherence, entanglement, and superposition. This generates a unique bitstream for each quantum device as a fingerprint. The proposed QPUF design is evaluated on IBM and Google quantum systems and simulators. The deployment on the IBM quantum simulator (ibmq\_qasm\_simulator) has achieved an average Hamming distance of 50.07\%, 51\% randomness, and 86\% of the keys showing 100\% reliability.
	\end{abstract}

	\keywords{Cyber-Physical Systems (CPS)\and Smart Grid \and Cybersecurity \and System Security \and Security-by-Design (SbD) \and Physical Unclonable Functions (PUF)\and Quantum Physical Unclonable Functions (QPUF)}
	
	\section{Introduction}
\begin{figure}[htbp]	
		\centering
		\includegraphics[width=0.8\textwidth]{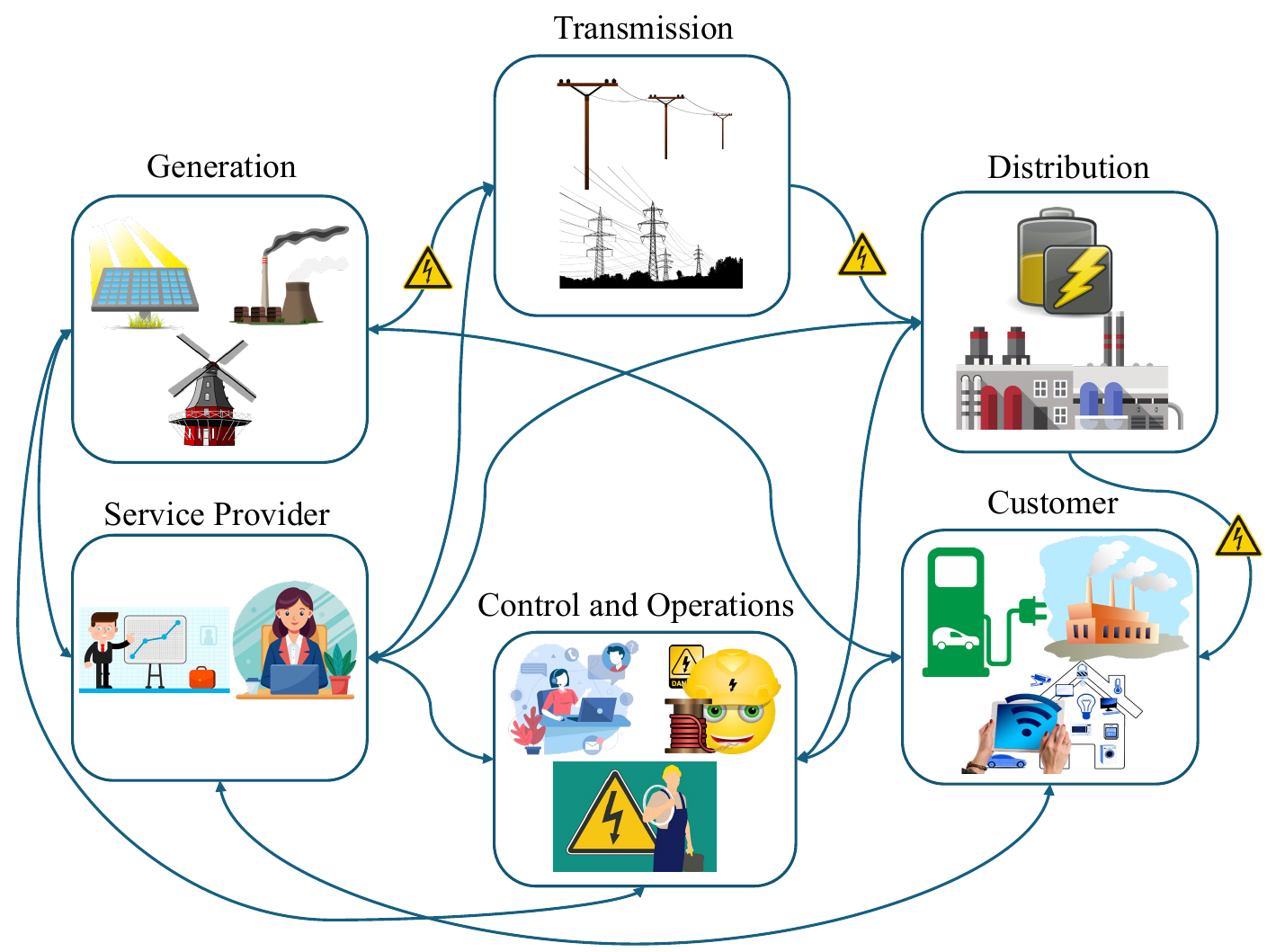}
		\caption{NIST Smart Grid Conceptual Model \cite{NIST_SmartGridFramework}}
		\label{fig:Output_1}	
	\end{figure}

	The advancement of electrical grids is required for enhanced power quality management, outage control, customer demand forecast, and power supply. The Smart Grid evolves from the technological integration of various state-of-the-art technologies that automate the electrical distribution processes. The smart integration of electronic devices provides real time data sensing and processing capabilities that help monitor outages, power turbulence, and metering infrastructure \cite{Patil2022}. Technological integration improves efficiency and increases reliability.  The Internet-of-Things (IoT) offers communication, control, and computation capabilities to the smart grid and enhances the communication and data processing flow in grid operations \cite{RadoglouGrammatikis2019}. The Smart Grid conceptual model by the National Institute of Standards and Technology (NIST) is depicted in Fig. \ref{fig:Output_1}.
	Along with the automation and advancement of energy infrastructure, generating and distributing renewable energy from various sources and their management is a challenge. Optimizing the grid functions and managing the distribution of resources requires a comprehensive approach. SCADA enables the management of generation, transmission, distribution, and control of energy in a Smart Grid. Its functionalities include managing grid operations, power quality, outage control, real-time demand analysis, and customer management \cite{cardwell2016efficacy}. 
	The salient features of a SCADA-enabled smart grid are:
	\begin{itemize}
		\item An efficient power demand response and management.
		\item Improved power quality with real-time power metric monitoring and control.
		\item A sustainable renewal energy resources distribution and management. 
		\item A simple and lightweight bi-directional communication framework among various entities in the energy distribution and control framework.
		\item A sustainable approach with reduced usage of greenhouse gas emissions.
		\item An efficient electricity trading approach by analyzing energy usage forecast and the microgrid's distribution efficiency. 
		\item A simple power outage control and self-healing capability with the integration of protective relays.
		\item An intelligent energy metering infrastructure with sustainable customer data protection and privacy-enhanced communication flow.
	\end{itemize} 
	
	SACDA enables secure analysis and control of communication in grid infrastructure. It consists of Remote Terminal Units (RTU) which monitors various energy generation metrics through electronic devices at the base station and control the electricity transmission process. On-field electronic devices perform data sensing of various electric parameters and relay data to the control center. The control center then initiates decision making and actuation processes ensuring quality power supply and distribution \cite{Kumar2023}. The heterogeneous nature of IoT devices and their functionalities in a smart grid can bring in security vulnerabilities due to the bi-directional communication flow, heterogeneous electronic devices, and customers' personally identifiable information. 
	
	Security-by-Design (SbD) emphasizes the security of a system from the design or manufacturing stage and enhances the trust and confidentiality of the device and data as a default functionality. 
	Prominent SbD primitives include PUF, and Trusted Platform Module (TPM) and offer resource efficient security to smart electronic devices. The PUF-based approach provides a unique digital fingerprint for a smart electronic device using its inherent properties and offers reliable and tamper-proof security. The growth of semiconductor technology and the increasing market of IoT devices is estimated to reach trillions by 2030 making PUF an efficient security primitive for sustainable application of IoT devices. Quantum computing can enhance the information processing capability using the principles of quantum mechanics. This research work presents a novel concept of Quantum SbD for enhanced security and trust in quantum computing applications based on the principles of quantum mechanics. The proposed QSbD approach includes a novel approach for extracting a PUF digital fingerprint based on quantum superposition, entanglement, and decoherence principles. The proposed QPUF 2.0 framework presents a unique approach for extracting a PUF-generated bitstream response from noisy quantum computers using their inherent process variations and securing the communication among various entities in smart grid infrastructure managed through SCADA. 

	The rest of this paper is organized as follows. The conceptual overview of a smart grid is presented in section \ref{sec:grid}. Section \ref{sec:qc} discusses the conceptual background of quantum computing. Section \ref{sec:related} discusses related research on Quantum SbD and smart grid cybersecurity. Novel contributions of the proposed work are presented in section \ref{sec:Novel Contributions}. The architectural overview of the proposed QPUF 2.0 is discussed in section \ref{sec:QPUF}. The proposed QSbD framework QPUF 2.0 is discussed in section \ref{sec:QPUF-2}. QPUF Experimental validation results are presented in section \ref{sec:ExperimentalResults} and finally, the conclusion and
	future research directions are discussed in Section \ref{sec:conclusion}.

	\section{Overview of Smart Grid}
	\label{sec:grid}
	In this section, a conceptual overview of the smart grid is presented and its communication framework as defined by the National Institute of Standards and Technology (NIST) is discussed. \cite{NIST_SmartGridFramework}. 
	
	The smart grid can be defined as an automated control grid's operations and communication with self-healing capabilities and effective management of distributed energy resources. The term smart grid is coined to define a technologically enhanced and efficient electrical power grid with robust control and automation of energy generation, transmission, and distribution \cite{fang2011smart}. The complexity involved in conventional power grids where a top-down approach having a one-directional power flow from the generation subsystem to the consumer subsystem has proven to be inefficient with increasing electricity usage demand and renewable energy resource integration and management. The bi-directional information flow increases efficiency and improves energy resource management with an advanced communication infrastructure that includes protective relays, IEDs, and circuit breakers performing data sensing, actuation, and processing in real-time \cite{Gaspar2023}. 
	
	Smart Substations (SS) play an important role in the electrical distribution framework. The main functionalities include voltage control, equipment monitoring, and fail-safe protection. Substations operate at AC/DC voltage and have a step-up transformer to increase the voltage to transfer power over long distances. While a step-down transformer enables lowering the voltage for efficient compliance with the electrical power voltage at the customer. An SS enables fail-safe protection by monitoring the electrical system power plow, identifying any malfunctioning equipment, and issues with voltage control through the advanced SCADA-based real-time monitoring and control through protective relays. Relays play an important role in grid infrastructure management. The functionalities include self-healing, monitoring equipment faults, and regulating voltage and current fluctuations \cite{Gaspar2023}. 
	\subsection{Supervisory Control and Data Acquisition (SCADA)}
	SCADA facilitates intelligent management and control of monitoring various critical infrastructures such as telecommunication, power plants, and industries. SCADA systems include smart electronic devices facilitating intelligent data sensing for monitoring various critical parameters which then relay data to a centralized control system for processing and analysis through Human Machine Interface (HMI). In electrical distribution and monitoring systems, the \textbf{Intelligent Electronic Devices} perform data sensing related to power metrics which helps in monitoring the power quality and its transmission. 
	
	IEDs along with \textbf{Phasor Measurement Units (PMU)} have various functionalities that include protective relaying, voltage and power frequency estimation at electrical distribution lines, power equipment data processing with microseconds resolution, and voltage control. These devices also include actuators that perform operations based on commands from \textbf{Remote Terminal Unit (RTU)} which is a gateway for controlling the sensors and actuators at a base station. IEDs are integrated microprocessor controllers for monitoring the power system equipment. IEDs consist of voltage, and current sensors to obtain power metric measurements, wired, wireless, and serial communication modules to interface with RTU, power supply inputs, onboard memory, and analog-to-digital converters \cite{Vaidya2022}.
	
\begin{figure*}[htbp]	
		\centering
		\includegraphics[width=1\textwidth]{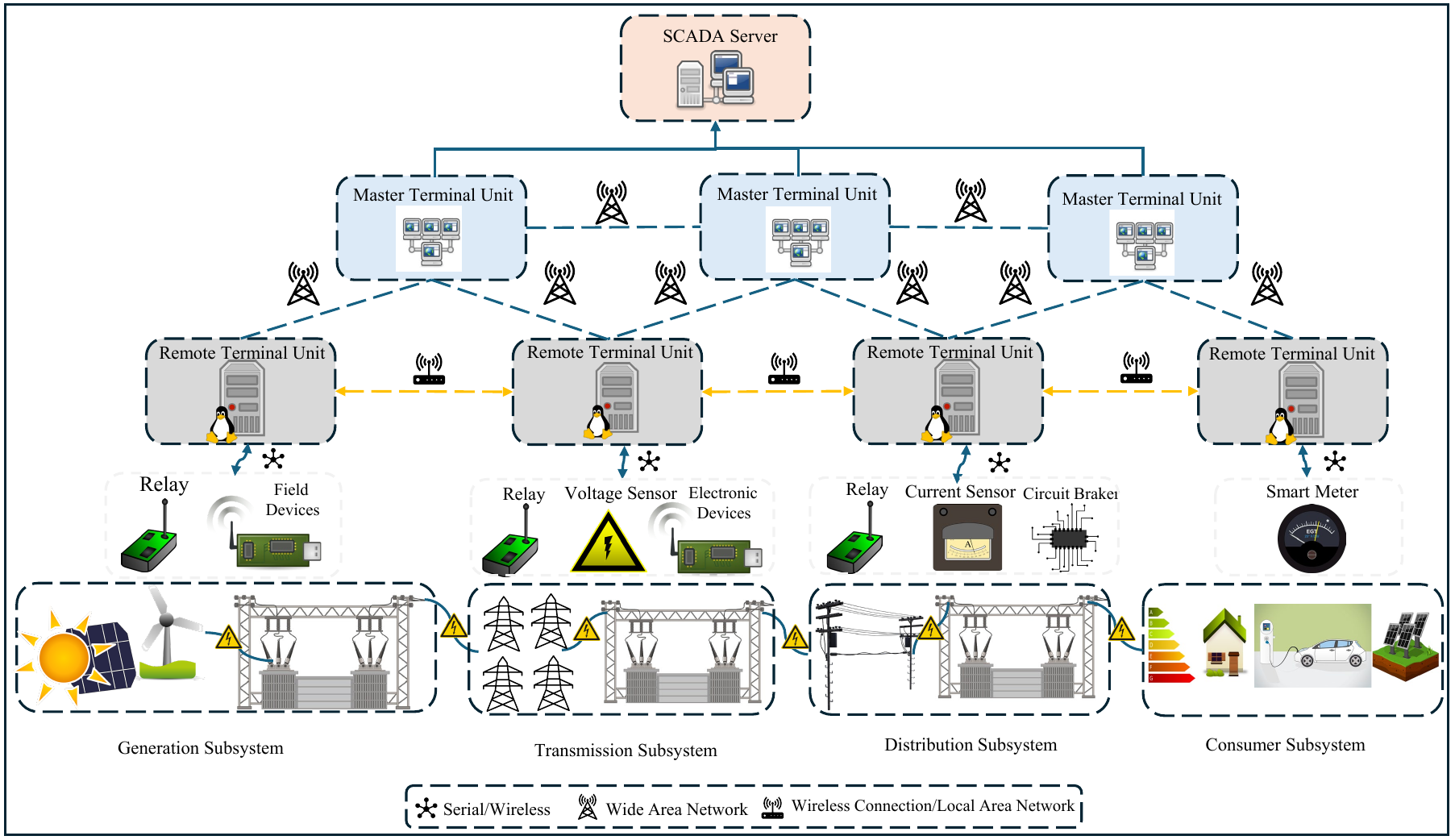}
		\caption{Architecture of Smart Grid}
		\label{fig:SCADA}	
	\end{figure*}

	RTUs have communication capabilities to connect with IEDs and relays at various subsystems enabling remote control, monitoring, and actuation. On top of RTUs, the \textbf{Master Terminal Units (MTUs)} provide high-level system control logic and communicate with RTUs and central command control for data analysis, processing, and storage through HMI \cite{Pliatsios, Gaspar2023}. The operator/command center is part of SCADA systems and supports HMI. Communication among various RTUs and MTUs is facilitated using wired and wireless technologies such as ModBus, DNP3, and optical fiber-based infrastructure. Programmable Logic Controllers (PLC) in SCADA are centralized electrical control systems that communicate with IEDs at the physical layer through RTUs \cite{Acharya2020}. Modern RTUs can communicate with IEDs located at substations through the RS-485 serial communication protocol. RTUs receive AC measurement inputs from voltage transformers, and current transformers and perform fault detection in real-time \cite{Sayed2017}. MTUs and RTUs work in a master/slave architecture where MTUs provide control logic and RTU controls IEDs. The communication and control of the power distribution framework in SCADA is presented in Fig. \ref{fig:SCADA}.
	\subsection{Smart Grid Infrastructure}
	The smart grid infrastructure has bidirectional information and power flow. Smart grid monitoring includes managing grid reliability, failure protection, and power equipment monitoring in its infrastructure which are essential for ensuring power quality.
	Electricity generation and transmission in a smart grid can be facilitated at various sources like smart farms with advanced solar cells, electricity trading from smart homes, and electric car charging stations. Various Blockchain-based solutions for energy trading have been explored where a smart and automated way of energy trading between a user at a smart home and grid can be facilitated in real-time. Similarly, electric vehicle-grid communication facilitates energy trading from a charging station when required to reduce the load on the grid. These approaches help in demand management during extreme weather conditions which can increase electricity usage. 
	
	The \textit{Microgrid} is an emerging smart grid subsystem with autonomy to island power generation, transmission, and distribution. The low-voltage electricity from various renewable energy resources from solar panels at home, and wind turbines at farms can be disconnected from the macro grid and can drive the energy requirements of individual consumers without relying on the macro grid. Microgrid requires efficient communication infrastructure with enhanced reliability and authentication protocols for ensuring the secure operation of microgrid.
	\subsection{Cybersecurity Issues in  Smart Grid}
	The data processing and communication flow in the energy distribution framework is facilitated using SCADA which has various entities with diverse communication and information processing capabilities. Providing a secure and efficient power supply framework without any security vulnerability is a challenging task due to the heterogeneity of various devices and their applications.
	
	\textit{Data Integrity Attacks} could arise when an adversary can intercept the communication between field sensors or IEDs and RTUs and possibly gain access to these devices and perform sniffing and snooping attacks. The Modbus communication protocol is used for sensor data transmission between RTUs and field sensors. The lack of standard benchmarks for security metrics associated with this protocol could jeopardize the communication framework of these devices. The SCADA's control unit receiving the sensor data cannot validate the trustworthiness of the data or verify the identity of IEDs which validates the requirement for a sensor integrity verification scheme \cite{GomezRivera2021}.
	
	\textit{Availability} of power system resources and their reliable operations is facilitated through effective control and management of protective relays and smart electronic devices monitoring power metrics. Unauthorized access and control through spoofing and repudiation can reduce the trustworthiness of smart grid operations where an adversary can replace a fake node with a reliable node and obtain access to communication. 
	
	\textit{Privacy} of consumer information and its authorized secure access is a prime requirement of advanced metering infrastructure. An adversary can extract power consumption analysis data from smart meters which can be vulnerable to various cyber-attacks due to their open working environment and can pose a threat to individual consumer privacy \cite{Akkad2023}.
	\subsection{Quantum Security-by-Design (QSbD)}
	SbD focuses on a sustainable consumer-level electronic system with security and privacy built as primitives at the design or entry level. 
	Adopting security practices to address the vulnerabilities and mitigate threats in smart electronics have proven to be inefficient due to limited processing and power capabilities of smart electronic devices. Developing applications with built in security practices is an effective way to mitigate threats. IoT applications require sustainable security both at the device and data level to counter any adversarial access that can jeopardize the integrity and authenticity of data \cite{Mohanty2020}. Security/Privacy by Design emphasizes the design and development stages combined with performing security risk analysis at various levels of application deployment and adopting security practices ensuring sustainable security with user centric approach. Facilitating security/privacy as proactive approaches can improve the resiliency instead of retrofitting these primitives as reactive solutions for various security threats and vulnerabilities in the emerging Internet-of-Everything (IoE) era \cite{Venkata2023}. 
	
	SbD works based on seven fundamental principles: 1) SbD approach should be proactive 2) Security/Privacy should be a default primitive at the design or system development stage 3) SbD should be completely embedded in the architecture enhancing resiliency of the system considering security risks 4) Facilitating end-end security 5) Full functionality with enhanced performance 6) User-centric 7) Visibility \& Transparency \cite{Pescador2022}.
	
	This research explores the potential of Quantum Security-by-Design (QSbD), an emerging approach for robust security in high-performance computing applications. QSbD focuses on enhancing the security of Artificial Intelligence, Blockchain, and Machine Learning applications using the principles of quantum mechanics in the emerging quantum computing driven IoT era. This motivational idea of this research work can be a suitable approach for trusted authentication and secure communication in E-CPS driven by QSbD. QSbD is based on emerging Quantum Physical Unclonable Functions, Quantum Blockchain, and Quantum machine learning for improved device, and data security, and privacy in quantum computing. The principles of SbD can be adopted in QSbD as well with a focus on security and privacy as default features for a quantum computing application. The working principles of QSbD can be:
	\begin{itemize}
		\item A secure communication framework through quantum teleportation for trusted data transfer
		\item A mutual device attestation technique using state-of-art quantum cryptography based on the principle of quantum state unclonability.
		\item A secure unique device identity attestation functionality through quantum device fingerprinting using process variations.
		\item Information processing, storage, and attestation capabilities secured through the quantum digital signature mechanism. 
	\end{itemize}
	
	\subsection{Physical Unclonable Functions} 
	
	SbD using hardware focuses on embedding security as a functionality based on the hardware characteristics and properties. The increasing demand and market for smart electronic devices validate the requirement for the security/privacy of a system based on the hardware with minimal performance trade-offs. Physical Unclonable Functions (PUF) and Trusted Platform Modules (TPM) are the most widely used state-of-art security primitives based on hardware. PUF can enhance the trust and integrity of an electronic device by enabling secure digital fingerprint generation using hardware properties based on Integrated circuit manufacturing variations (IC). A PUF is built based on IC technology that maps micro-manufacturing process variations to a unique cryptographic key. The unique PUF-generated identity cannot be tampered with or can be regenerated on other hardware \cite{Chanda2022} thereby making it a robust hardware-assisted security primitive. 
	
	A PUF has a challenge as input and response as output where a random function maps the intrinsic hardware properties as a challenge input and generates a unique bitstream as response. A PUF on a chip when tested with two challenge inputs $C_{i}$, $C_{i+1}$ will generate unique responses $R_{i}$, and $R_{i+1}$ respectively. The obtained responses will be unique since each challenge will work differently on underlying hardware resulting in unique responses ($R_{i}$$\neq$$R_{i+1}$). This reliability and tamper-proof hardware security primitive can also generate unique responses on different hardware with similar PUFs due to varying process variations during chip fabrication in IC development. The PUF module on device $d_{1}$ generating $R_{1}$ from $C_{1}$ will not be the same as $R_{2}$ generated from the same challenge $C_{1}$ on device $d_{2}$. 
	\section{Quantum Computing Background}
	\label{sec:qc}
	This section provides a brief overview of quantum computing concepts, logic gates, hardware resources, and information processing capabilities.
	\subsection{Qubit Overview}
	A Qubit is the basic unit of quantum information. Bits can exist only in one of the two states at a time whereas a qubit can exist in the superposition of both 0 and 1 states at the same time \cite{Bhat2022}. In comparison to classical bits, qubits will have outputs with an equal probability of being 0 or 1 and are represented as follows:
	$\ket{0} = \begin{bmatrix}1\\0 \end{bmatrix}$,
	$\ket{1} = \begin{bmatrix}0\\1 \end{bmatrix}$\\
	The quantum state of a qubit is represented as  $\begin{bmatrix}X\\Y \end{bmatrix}$. When the quantum state of a qubit is measured, $X ^{2}$ is the probability of obtaining 0 state and $Y^{2}$ is the probability of attaining state 1. Under normalization conditions, the squares of amplitudes of probabilities will be equal to 1.
	\begin{center}
		$|X|^{2}$+$|Y|^{2}$=1.\\
	\end{center}
	
	The amount of information that can be represented by qubits is approximately half the number of classical bits required to represent the same information \cite{Sridhar2023}. Various quantum hardware with diverse physical qubit control architectures is illustrated below: \\
	
	\textit{Superconducting Qubits} are most commonly used in IBM quantum systems. The underlying hardware development includes the lithography process to pattern superconducting circuits made of niobium and depositing these materials on a silicon substrate. Qubit structures are created based on Josephson junctions that control the quantum state of qubits and their behavior by varying electromagnetic fields. These qubits work based on the principle of superconductivity that defines a material working at extremely low temperatures and conducts current with zero resistance.\\
	\textit{Trapped Ion Qubits} work based on the internal energy of trapped ions to store, manipulate, and control the information processed by the qubits. IonQ uses trapped ion qubits in their quantum computing systems. Through a laser-based control on the ions, the qubits encoded at different energy levels inside an ion can be controlled.\\
	\textit{Topological Qubits} are built based on the topological states of quantum mechanics to achieve noise-free qubits. These qubits are used in Microsoft quantum systems and are controlled by manipulating their structure using chemical bonds.\\
	\textit{Quantum Dots:} Quantum Dots are tiny-sized semiconductor particles that carry electricity and their energy levels are defined by the laws of quantum mechanics. These are tiny semiconductor particles where electrons are tightly packed, leading to the formation of energy levels. These energy levels can be manipulated through light or electricity.
	\subsection{Quantum Gates Overview}
\begin{figure}[htbp]	
		\centering
		\includegraphics[width=1\textwidth]{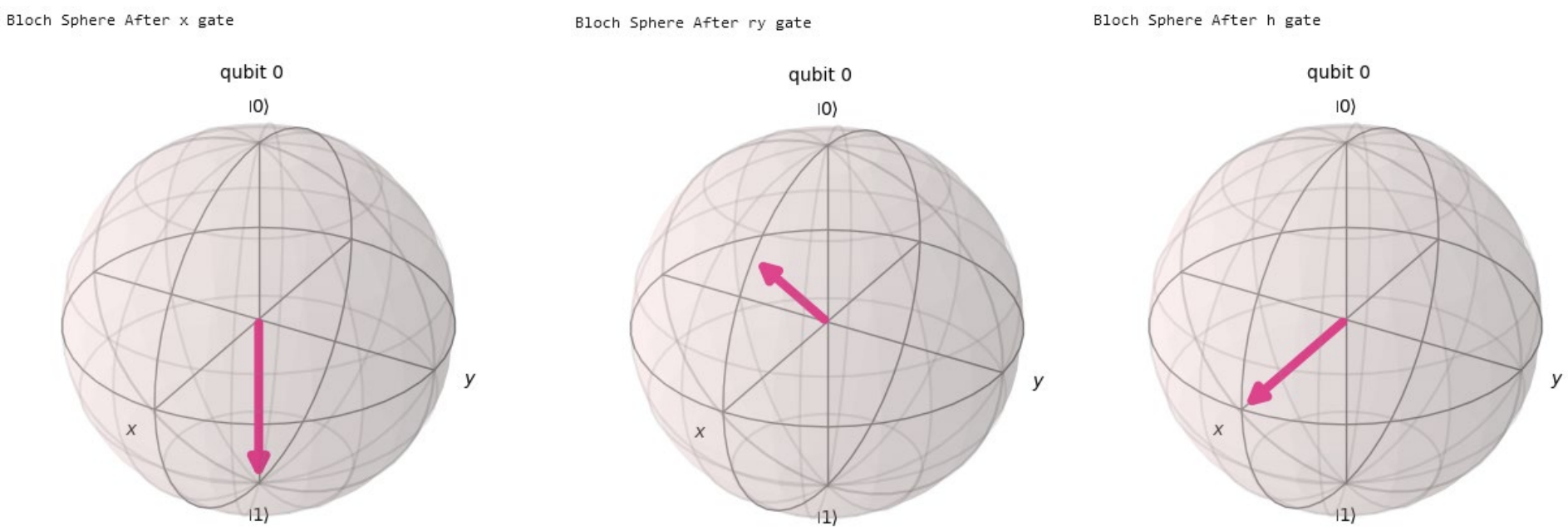}
		\caption{Bloch Sphere Representation of Quantum State Variations of a Qubit}
		\label{fig:bloch}
	\end{figure}

	Quantum gates are building blocks of quantum circuits and algorithms that support quantum state manipulation using various operators. In digital logic, AND, OR, NOT, and XOR gates work on bits and process the information. Analogous to digital logic, quantum logic gates work on quantum bits whose information state is a superposition of two states represented by the vectors \cite{Singh2021, Ardelean2022}. The quantum gates are either single or two qubit gates and are applied to qubits to manipulate their quantum state. When the state of a qubit is measured, it collapses to one of the two states: either 0 or 1. Fig. \ref{fig:bloch} shows the geometrical representation of qubit's state changes when various quantum gates are applied \cite{Shaik}.
	
	Various single and two qubit logic gates are:
	
	\textit{Pauli X-Gate:} The Pauli X-gate  flips the quantum state of a qubit. If the quantum state is 1, it flips the state to 0 and vice versa. Its functionality is like a NOT gate in digital logic.\\
	
	\textit{Identity Gate:} The I-gate is similar to an identity gate or buffer and it does not manipulate the qubit state and can be used to retain the quantum state. This gate can introduce decoherence in the quantum state of a qubit by introducing the delay.
	
	\begin{center}
		$I gate\ket{0} = \begin{bmatrix}1 & 0\\0 &1 \end{bmatrix}\begin{bmatrix}1\\0 \end{bmatrix}=\ket{0}$,

	\end{center}
	\begin{center}
		$I gate\ket{1} = \begin{bmatrix}1 & 0\\0 &1 \end{bmatrix}\begin{bmatrix}0\\1 \end{bmatrix}=\ket{1}$
	\end{center}
	\textit{Hadamard Gate:} The Hadamard gate is a single qubit quantum gate that manipulates the state of a qubit and places it in superposition. When the state of the super-positioned qubit is measured, it falls to either one of the two possible states which are 0 or 1. 
	\begin{center}
		$H Gate = (1/\sqrt{2})\begin{bmatrix}1 & 1\\1 &-1\end{bmatrix}$\\
	\end{center}
	\textit{RY Gate:} The RY gate is a single qubit tunable rotational quantum gate that performs qubit's state rotation around the y-axis of the Bloch Sphere. RY gate is used to place the qubit in an unknown quantum state which can introduce more unpredictability. 
	
	\textit{CNOT Gate}: The Controlled NOT gate is a two-qubit quantum gate that binds the quantum states of two qubits in such a way that the quantum state change of one qubit affects the other qubit. To realize CNOT, the target qubit will be driven by the control qubit's resonance frequency at the hardware level. Logically, the quantum state of the target qubit is flipped using X-gate if the control qubit is in $\ket{1}$ state.

	\section{Related Prior Works}
	\label{sec:related}
	This section gives a comprehensive overview of state-of-the-art research on Quantum PUF and smart grid cybersecurity. Also, a comprehensive review of QPUF and smart grid SbD solutions are presented in Tables \ref{Table:Research-1}, and \ref{Table:Research-2}.
	
	A Quantum hardware verification approach using qubits' cross talk was proposed in \cite{Chwa2023} using a QPUF which works on superconducting transmon qubits from IBM. Their work evaluated a QPUF design using a simple Hadamard gate application using microwave pulse on control qubit and its impact on target qubit due to cross talk was explored. QPUF designs based on quantum logic gates using quantum's superposition and decoherence phenomena were presented in \cite{Phalak2021}. Their work proposed a Hadamard gate based PUF to explore superposition phenomena and an idle gate driven design to study the impact of decoherence for realizing a unique bitstream as a PUF response. In comparison, this work proposed a novel architecture in \cite{bathalapalli2023} based on Hadamard gate PUF proposed in the above work \cite{Phalak2021} that supports challenge-response pair generation and PUF metric evaluation for various quantum systems from IBM. 
	
	A Quantum Readout (QR) PUF was proposed in \cite{cryptoeprint:2009/369} which works on classical PUF. The PUF design is tested with challenges and responses in quantum states. The QR PUF is claimed to be more effective than a classical PUF as it is based on the no-cloning principle which states that it is impossible to duplicate or clone the unknown quantum state of a qubit. A simple PUF-based key exchange protocol based on the quantum physical unclonability principle was presented in \cite{Boris2012} which proposes a PUF-based quantum BB84 key exchange protocol with CRP being converted to qubits. A QPUF multi-factor authentication algorithm based on the principle of no-cloning theorem was proposed in \cite{Galetsky2022}. This work also includes an enrollment authentication mechanism through QPUF using QTOKSim, a quantum token-based authentication simulator.
	
	A novel Quantum tunneling PUF titled Neo PUF has been proposed in \cite{Chuang2021} which works by storing the PUF signature inside ultra-thin oxide. This PUF works based on manufacturing variations in oxide. 
	\begin{table*}[htbp]
		\centering
		\caption{Related Research on Quantum PUF}
		\label{Table:Research-1}
		\begin{tabular}{|m{2.1cm}| m{2.8 cm}|m{1.75 cm}| m{2.3 cm}|m{2.8 cm}| m{2.0cm}|}
			\hline
			\textbf{ Work}&\textbf{Quantum Hardware}&\textbf{Logic Gates}&\textbf{Quantum  Principles}&\textbf{QPUF Metrics}&\textbf{Challenges}\\ 
			\hline
			\hline
			Phalak, et al. (2021)\cite{Phalak2021} & ibmq\_essex and ibmq\_london& H, Ry, Measurement, and I & Superposition and Decoherence & intra-HD ibmq\_essex (13.82\%) and ibmq\_london(3.94\%) & No CRP\\
			\hline	
			Chwa, et al. (2023)\cite{Chwa2023} &  IBM Falcon r5.10 and r5.11 27, ibm cairo, ibm hanoi, ibmq mumbai, and ibmq kolkata & Hadamard & Quantum Crosstalk & NA & No impact of Control on Target \\
			\hline	
			Bathalapalli, et al. (2023)\cite{bathalapalli2023} & ibmq\_belem, ibmq\_lima, and ibmq\_quito & Hadamard, Ry & Superposition &40\% Intra-HD ibmq\_lima, 25\% Uniqueness &Less uniqueness and Diffuseness\\
			\hline
			Khan, et al. (2023)\cite{Khan2023} & 5-qubit processors & Ry and Rx gates & Entanglement & 1 state probability- q[0]-76\%, q[1]-87\% & No evaluation of Intra and Inter HD\\ 
			\hline
			\textbf{QPUF 2.0 (Current Paper)} & ibm\_osaka, ibm\_kyoto, and ibm\_sherbrooke & H, Ry, CNOT, and I-gate & Quantum Entanglement, Decoherence, and superposition &50\% HD, 51\% Randomness & Can improve uniqueness\\
			\hline
		\end{tabular}
	\end{table*}	
	A novel Quantum Key Distribution (QKD) protocol for secure communication in Quantum computing applications was proposed in \cite{Padamvathi2016}. QKD protocol works based on the principle of Heisenberg's uncertainty principle. In this protocol, an unverified party can't intercept the communication on the Quantum channel between two trusted entities \cite{Rahman2019, Sangari2023}. In comparison to the works cited above, this work experimentally validates the QPUF design implemented using quantum Logic gates and clearly defines the PUF signature generation process based on quantum mechanics principles that differentiate each quantum hardware.
	
	For smart grid security and countering man-in-the-middle attacks, Quantum-Sim, a secure smart grid communication framework based on Quantum key distribution (BB84) protocol is proposed in \cite{Lardier2019}. A secure IoT device attestation framework using QPUF was presented in \cite{Khan2023}. Their work proposed a QPUF mechanism based on the principles of quantum mechanics and the BB84 protocol. 
	
	A client’s server handshaking protocol is validated with noisy quantum computers using Hadamard gate based QPUF design.  A unique approach to fingerprint quantum servers executing quantum circuits from users based on the error rates of various quantum gates on different quantum devices is proposed in \cite{Wu2024}. Using randomized benchmarking, error rates of various gates in different circuits with and without an identity gate and their probability distributions of states are determined and evaluated in this work to identify the server. 
	
	A PUF-based approach for authenticating quantum computers is proposed in \cite{Morris2023}. SRAM PUF module has been used for fingerprinting quantum computers. This work is mainly based on implementing SRAM PUF at cryogenic temperatures. This work also examines the feasibility of SRAM PUFs in Quantum computers using liquid nitrogen to cool SRAM memories at -195 degrees Celsius. This work, however, does not propose a quantum hardware based QPUF and works on SRAM PUF which has a limited number of CRPs.
	
	A PUF-based mutual authentication scheme for Vehicle to Grid (V2G) proposed in \cite{Sharma2021} focuses on secure communication among Vehicles, Charging Stations (CS), and Grid Servers.  Bi-directional communication exists between CS and GS where CS communicates location, charging rate, duration, and energy demand forecast details with GS. Similarly, the EV shares the location details with GS for identity and attestation. For mutual authentication and security of data, an efficient and secure authentication framework (ESAF) based on PUF was proposed in \cite{Mehta2024}. A lightweight PUF-based protocol for effective communication between smart meters and neighborhood gateways by embedding a PUF with each smart meter device was proposed in \cite{Kaveh2020}. A PUF-based authentication protocol for securing communication between various substations and control centers in E-CPS was presented in \cite{Hutto2022} which proposes a PUF-based approach for securing IEDs with minimal overhead. Their work also considered an attack scenario that includes a substation with a fake IED. 
	
	SRAM PUF-based secure RTU and IED communication framework with an attestation mechanism for IEDs using SRAM PUF in \cite{GomezRivera2020} claims to ensure the data integrity from IEDs communicating with an RTU through a robust authentication framework. An authentication protocol for SCADA-enabled systems for Industry 4.0 was proposed in \cite{Rivera2021}. Their work envisions data flow integrity and non-repudiation by using SRAM PUF for authentication due to high entropy and hardware-generated randomness. In \cite{Jadhav}, $IED_{PUF}$ probe-based IED authentication mechanism which is a secure hardware-software solution is proposed. In \cite{GomezRivera2021}, a Blockchain-integrated PUF framework for sensor authentication in the SCADA framework is proposed. The authors validated a smart contract based PUF CRP enrollment and authentication mechanism. Their work claims to address a major security issue through modeling attacks on PUF CRPs predicting PUF responses. In comparison to the above cited works on SbD in E-CPS, the proposed QPUF 2.0 presents a quantum hardware attestation framework through QPUF for the security of on-field RTUs and MTUs performing data processing. 
	
	\begin{table*}[htbp]
		\centering
		\caption{Related Research on Smart Grid Cybersecurity}
		\label{Table:Research-2}
		\begin{tabular}{|m{3cm}|m{3 cm}|m{2 cm}| m{3.2 cm}| m{3 cm}|}
			\hline
			\textbf{ Work}&\textbf{Approach}&\textbf{Hardware}&\textbf{Security Primitive}&\textbf{Possible Challenges}\\ 
			\hline
			\hline
			Hutto, et al. (2022) \cite{Hutto2022}& IED Attestation  & IC & SRAM PUF& Requires Database\\
			\hline	
			Sharma, et al. (2021)  \cite{Sharma2021} & PUF-V2G & NA & PUF-based Vehicle-to-Grid & NA\\ 
			\hline
			Gomez Rivera, et al.(2020) \cite{GomezRivera2020} & SPAI & IC & SRAM PUF (RTU-PLC)& PUF-AES Hardware Overhead\\
			\hline
			Vaidya, et al. (2022) \cite{Vaidya2022} & IED Attestation &Analog-Digital circuit (ADC) & PUF-ADC &  Hardware Overhead\\
			\hline
			Cao, et al. (2021)\cite{Cao2021} & Authenticated Metering Infrastructure &   NA & PUF-Smart Meter &  No Hardware\\
			\hline
			Gomez Rivera, et al. (2021) \cite{GomezRivera2021} & PUF-Blockchain for SCADA & SRAM PUF &Smart Contract &  Scalability\\
			\hline
			\textbf{	Current Paper} & QPUF for IED Attestation  & Quantum Computers& Hadamard, RY, CNOT& NA\\		
			\hline
		\end{tabular}
	\end{table*}

	\section{Contributions of the Current Paper}
	\label{sec:Novel Contributions}
	\subsection{Research Questions Addressed in the Current Paper}
	The research questions addressed through this work are:
	\begin{itemize}
		\item How to ensure security and privacy in E-CPS using quantum computing?
		\item How quantum computing can be feasible for improving the smart grid's operational workflow?
		\item How to provide security and privacy to the device and data in a SCADA-driven smart grid infrastructure with heterogeneous functionalities?
		\item How to ensure the reliability of a QSbD solution due to the inherent noise in quantum systems and the instability in a qubit's quantum state? 
	\end{itemize}
	\subsection{Challenges in Solving the Problem}
	The challenges involved in this research are:
	\begin{itemize}
		\item Developing a standard security solution for the integrity and authenticity of various entities in the energy distribution cycle.
		\item The scalability issues associated with the integration of billions of smart electronic devices or IEDs in SCADA-Smart Grid with noisy quantum computers.
		\item Exploring digital fingerprint generation schemes like PUF on noisy quantum systems can be challenging due to the environmental impact which affects the QPUF's reliability in quantum cryptography.
		\item The access to quantum systems is still facilitated by very few companies through cloud-based access which is a challenge for the emerging Quantum Chain of Things (QCT or QIoT)
	\end{itemize} 
	\subsection{Novel Contributions of the Current Paper}
	The contributions of this research work are:
	\begin{itemize}
		
		\item A novel quantum-hardware assisted SbD framework for SCADA-driven electrical distribution framework.
		\item A Quantum PUF-based approach for the security of IEDs and protective relays and their communication framework using Quantum hardware.
		\item A QPUF design based on quantum mechanics principles of quantum decoherence, superposition, and entanglement.
		\item An experimentally evaluated QPUF design with reliability on noisy quantum computers.
		\item A strong QPUF design that generates higher cryptographic keys for enhanced security.
		
	\end{itemize}
	
	\section{QPUF 2.0: A Quantum Hardware PUF based on Quantum Decoherence and Entanglement}
	\label{sec:QPUF}
	This section discusses the proposed QPUF design and physical quantum hardware calibration details. Also, a detailed comparative analysis of PUF with QPUF is also presented.
	
\begin{figure}[htbp]	
		\centering
		\includegraphics[width=1\textwidth]{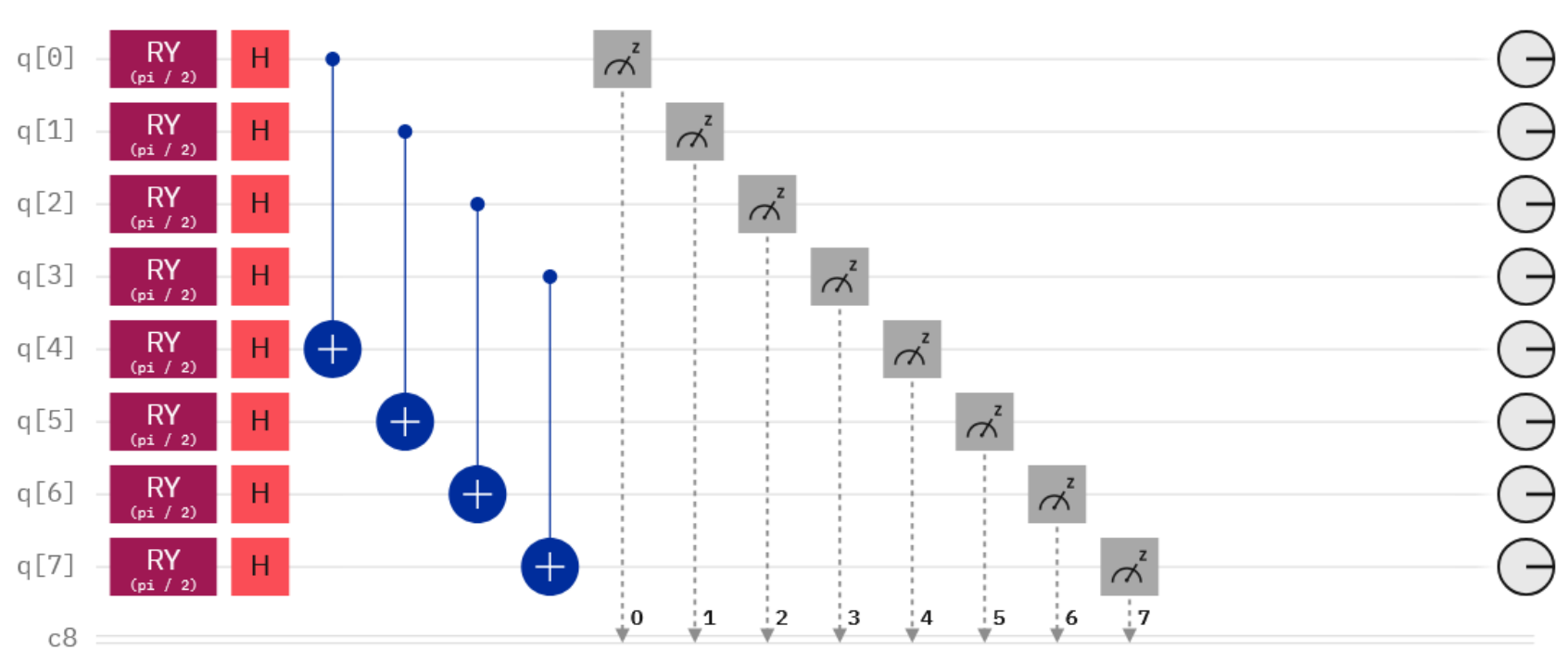}
		\caption{Proposed QPUF Design based on Quantum Logic Gates}
		\label{fig:QPUF_2}
	\end{figure}

	The proposed QPUF circuit is an 8-qubit architecture with single and two-qubit quantum gates: Ry gate, Hadamard gate, CNOT, and idle gates. I-gate is applied to retain the quantum state or qubit's coherence for a specified time. However, the physical realization of I-gate in superconducting transmon circuits introduces decoherence propelling the qubit to lose its quantum state or coherence. The main motivation for this research is to explore the decohering nature of superpositioned qubits to generate a unique bitstream as QPUF response from the quantum hardware. Also, the proposed architecture works on entangling a set of qubits to verify the impact of a qubit's decoherence during quantum entanglement. The architecture of the proposed QPUF design is shown in Fig. \ref{fig:QPUF_2}. 
	
	Initially, all the qubits are initialized randomly to either 0 or 1 state using Pauli X-gate. Ry gate, a single qubit gate is applied to all the qubits placing them in an unknown quantum state. This gate introduces the desired unpredictability for the qubit's quantum state in QPUF. To place the qubits in the superpositioned state which has an equal probability of obtaining the 0 and 1 states, the Hadamard gate is then applied to all the qubits. Controlled not gate (CNOT) is then applied to all the qubits that exclusively entangle the quantum state of the first four qubits with the last four in the circuit such that the quantum state variations of one qubit affect the other qubit. Once the quantum states of all qubits are entangled, Idle gates are applied to the control qubits. The physical realization of idle gates might introduce decoherence due to the qubit's interaction with the environment. The quantum state coherence of these superpositioned qubits with the delay is evaluated to generate a QPUF signature. This QPUF design explores the impact of the control qubit's decoherence on the target qubit. The state measurement of all the qubits can generate a unique bitstream as a QPUF response that can be used as a digital unclonable fingerprint for that quantum system. Detailed overview of accessing the quantum system and performing logic gate operations in the proposed QPUF 2.0 are presented in Algorithms \ref{algo:QPUF}, and \ref{algo:QPUF_2}.
	\subsection{Quantum PUF}
\begin{figure}[htbp]	
		\centering
		\includegraphics[width=0.85\textwidth]{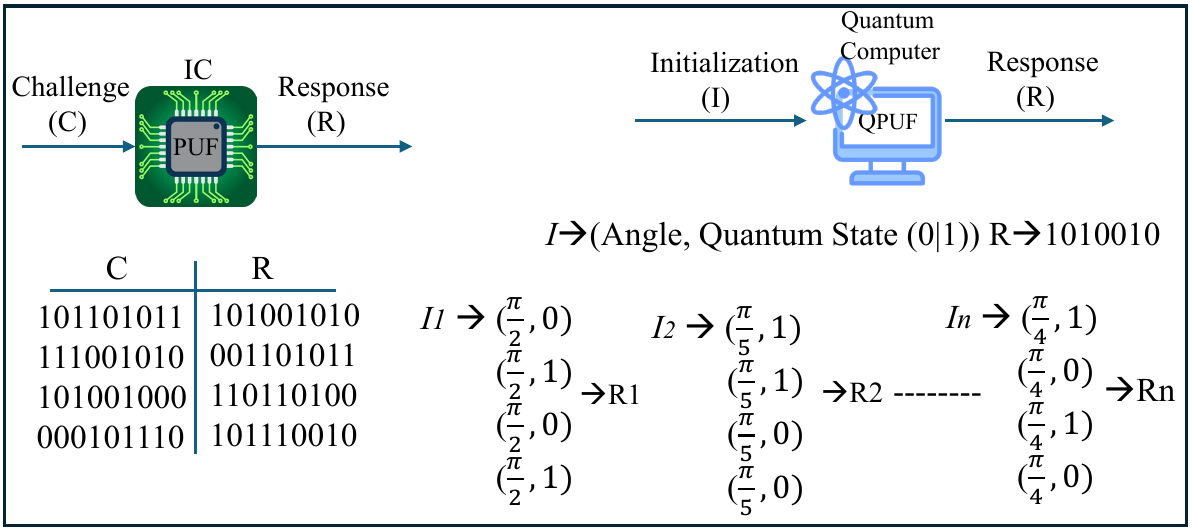}
		\caption{PUF vs QPUF}
		\label{fig:PUF-QPUF}	
	\end{figure}

	Quantum physical unclonable functions (QPUF) is a hardware cryptographic primitive for security and reliability in quantum computing applications. Since the underlying architecture and hardware of a quantum computing system are based on superconducting circuits, the PUF is assumed to have great significance in quantum computing applications. The CRPs in Quantum PUF are typically in unknown quantum states rather than binary states in the case of a PUF. The CRPs in a quantum PUF are the superpositioned quantum states with an equal probability of obtaining 0 or 1. QPUF works based on the principle of quantum mechanics which guides the working of quantum hardware with various quantum gates performing qubits' quantum state manipulation. It is practically infeasible to rebuild or emulate a Quantum PUF design and estimate the quantum state changes of qubits on another quantum hardware as defined by the principle of \textit{'no-cloning theorem and quantum physical unclonability'} \cite{Boris2012}. Fig. \ref{fig:PUF-QPUF} shows the difference between PUF and QPUF topologies.
	
	\subsection{Superconducting Quantum Circuits}
	Superconducting quantum circuits are the most widely chosen hardware for realizing quantum circuits. Given the scalability and ease of control through microwave operability, the realization of quantum hardware is becoming simpler and easier. Superconducting circuits operate at extremely low temperatures where the electrical resistance is absent and supports infinite conductivity. These circuits operate at extremely low temperatures, typically around -250 degrees Celsius. Superconducting circuits consist of Josephson junction, inductance, and capacitance elements. These circuits conduct at extremely low temperatures where the direction of the flowing current in the Josephson junction typically indicates the quantum state of a qubit. Mathematically, a qubit is represented as follows \cite{Bardin2020}:
	\begin{center}
		$\ket{q} = \begin{bmatrix}q0\\q1 \end{bmatrix}$$\rightarrow$ q0$\ket{0} + q1\ket{1}$,\\
		$p{\ket{0}} = {|qo|}^{2}$,\\
		$p{\ket{1}} = {|q1|}^{2}$
	\end{center}
	A Josephson junction, typically a building block of superconducting circuits consists of two superconducting electrodes sandwiched by a thin insulating barrier. Quantum state readout and manipulation are typically performed using microwave photons interacting with qubits. Typically, electrons form cooper pairs which can tunnel through the Josephson junctions based on the phase difference of the superconductors. The Josephson junctions create a harmonic energy level system for physical qubits to transition. Anharmonicity of energy states is ideal for superconducting transmon qubits realization \cite{Huang2020}. The nonlinear inductor or Josephson junction typically leads to a non-equidistant two-energy level system where the ground state typically represents quantum state $\ket{0}$ and the excited state is represented by $\ket{1}$. Physically, the gate operations on qubits are realized using microwave pulses at a specified phase, amplitude, and frequency in resonance with qubit's frequency to transition from ground to excited state \cite{Kjaergaard2020}.
	\subsection{Unclonable Characteristics of Quantum Hardware}
\begin{figure*}[htbp]
		\centering
		\begin{subfigure}[b]{0.4\textwidth}
			\centering
			\includegraphics[width=0.75\textwidth]{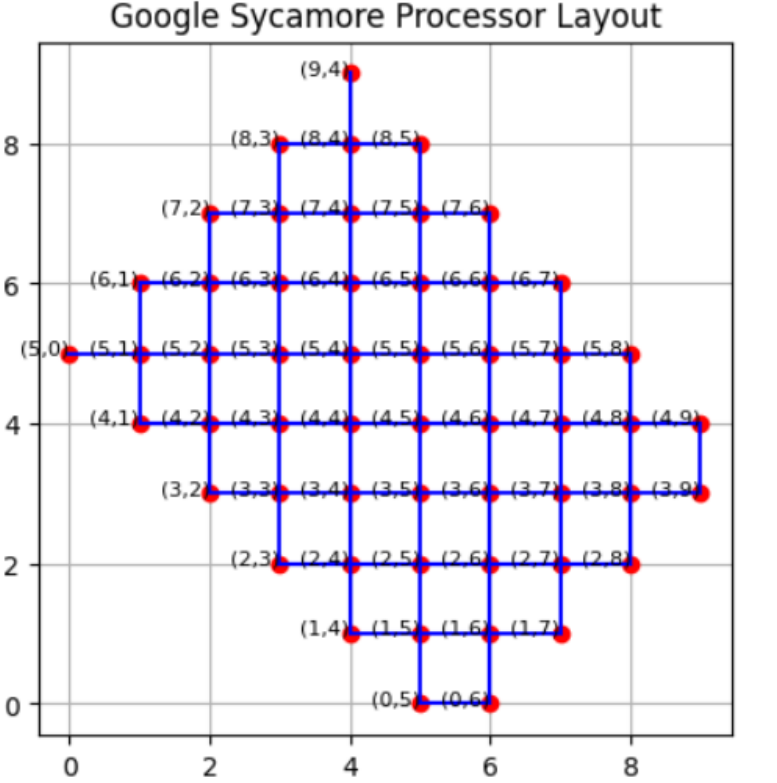}
			\caption{Physical Qubit Layout of Google Quantum Sycamore Processor}
		\end{subfigure}
		~
		\begin{subfigure}[b]{0.55\textwidth}
			\centering
			\includegraphics[width=1\textwidth]{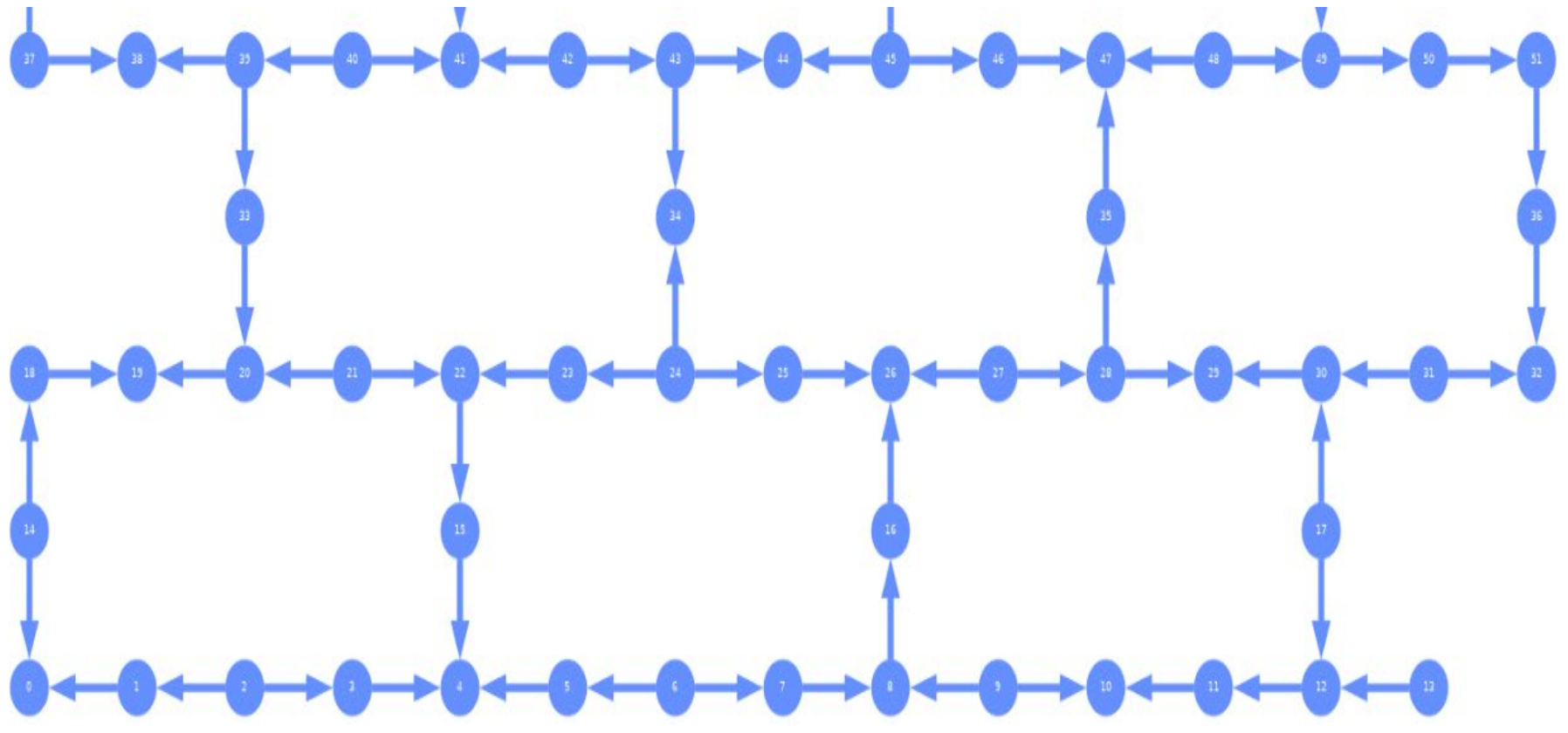}
			\caption{Physical Qubit Layout of IBM's Quantum System ibm\_osaka \cite{ibm_quantum_access}}
		\end{subfigure}
		\caption{Physical Qubit Mapping of Quantum Systems}
		
		\label{fig:map}
	\end{figure*}

	Quantum circuits are subjected to process variations during manufacturing processes such as fabrication, Josephson junction placement, lithography, and base metal placement. The environmental impact on these superconducting circuits which work at extremely low temperatures causes the physical qubits to experience decoherence and lose their quantum state. Physical qubits' quantum state is practically evaluated using the current flowing through the Josephson junctions in superconducting transmon circuits. Depending on the direction of the current as either clockwise or anticlockwise, the quantum state of the qubit is defined as either 0 or 1 respectively \cite{Smith2023}. The physical qubit layout of various quantum hardware is shown in Fig. \ref{fig:map}.
	
	Physically, quantum logic gate operations are realized using microwave pulses interacting with these qubits at a specified magnitude, shape, and direction \cite{Torosov2023}. This change in microwave pulses is to obtain the desired change in the quantum state, a transition between energy levels based on an anharmonic oscillator that corresponds to the mathematical representation of the qubit's rotation around the x, y, and z-axis of the Bloch sphere \cite{Khan2023}.  Qubits' quantum state can also change due to cross talk which arises due to coupling of qubits in the hardware. The application of microwave pulse on one qubit might also affect the neighboring qubit while realizing the gate operations \cite{Phalak2021}. The quantum decoherence of entangled qubits is based on the alignment of qubits physically in the quantum hardware. The quantum entanglement of neighboring qubits and the decoherence of control qubits can decohere target qubits faster than the decoherence of distantly entangled qubits.
	\begin{algorithm}[ht]
		\caption{Accessing Quantum System and Building Circuits}
		\label{algo:QPUF}
		\small
		\begin{algorithmic}[1]
			\renewcommand{\algorithmicrequire}{\textbf{Input:}}
			\renewcommand{\algorithmicensure}{\textbf{Output:}}
			\REQUIRE  IBM Quantum API Token
			\ENSURE   QPUF Design
			\STATE Access IBM Quantum Learning
			\STATE Obtain IBM Quantum API Token
			\\ \textit{User $\rightarrow$ API Token}
			\STATE Access Qiskit
			\STATE Build Quantum Circuit using Quantum Logic Gates
			\\ \textit{QPUF $\rightarrow$ Hadamard, Pauli-X,  Pauli-Y, CNOT,  Measurement Gates}
			\STATE Choose Quantum Backend
			\\ \textit{Quantum System $\rightarrow$ 127 Qubits }
			\\
			\begin{itemize}
				\item ibm\_kyoto, ibm\_osaka, ibm\_sherbrooke, ibm\_Brisbane
				\item Simulator $\rightarrow$ibmq\_qasm\_simulator
			\end{itemize}
			
			\STATE Execute the Circuit on the chosen backend and monitor the status of the job
			\\ \begin{itemize}
				\item Step 1: Job is being Validated,\\
				\item Step 2: Job is Queued, \\
				\item Step 3: Job is Actively Running,\\
				\item Step 4: Job has successfully Run
			\end{itemize}
			\STATE Obtain the result
			\STATE \textbf{Note: IBM Quantum Lab was sunset on May 15th, 2024}
		\end{algorithmic}
	\end{algorithm}   
	
	\begin{algorithm}[ht]
		\caption{QPUF Design Calibration}
		\label{algo:QPUF_2}
		\small
		\begin{algorithmic}[1]
			\renewcommand{\algorithmicrequire}{\textbf{Input:}}
			\renewcommand{\algorithmicensure}{\textbf{Output:}}
			\STATE Create a quantum circuit with 8 quantum and classical registers
			\STATE Initialize all the qubits q[0]-q[7] randomly with 0 or 1 state for each job
			\begin{itemize}
				\item  \textit{Choose X-Gate to initialize qubit to 1 state}\\
				\textit{State 1$\rightarrow$X[q]}
				\item  \textit{Each qubit is initialized to state 0 by default}
			\end{itemize}
			\STATE Place qubits in an unknown quantum state by applying Ry Gate
			\begin{itemize}
				\item Angle $\rightarrow$ 0-2 pi
			\end{itemize}
			\STATE Place qubits in superposition
			\begin{itemize}
				\item Apply Hadamard Gate\\
				\textit{	qubits' state probabilities $\rightarrow$ 50\%}
			\end{itemize}
			\STATE Perform quantum state entanglement
			\begin{itemize}
				\item Apply CNOT Gate\\
				\textit{q[0]$\rightarrow$ q[4]},
				\textit{q[1]$\rightarrow$ q[5]},
				\textit{q[2]$\rightarrow$ q[6]},
				\textit{q[3]$\rightarrow$ q[7]}
			\end{itemize}
			\STATE Control qubits by applying idle gates to evaluate the qubit's coherence with delay
			\begin{itemize}
				\STATE Apply I Gate\\	
				\textit{I[q[0]]},
				\textit{I[q[1]},
				\textit{I[q[2]},
				\textit{I[q[3]}
			\end{itemize}
			\STATE Perform quantum state measurement of qubits and store the result in a classical register corresponding to each qubit
		\end{algorithmic}
	\end{algorithm}

	\subsection{Physical Parameters of Quantum Circuits}
	QPUF signature is obtained by mapping the quantum mechanical qubit's properties to generate a unique bitstream as a signature. Quantum state manipulation of a qubit is performed at a microwave pulse level by sending a microwave pulse in resonance with the qubit's driving frequency. The qubit's T1 and T2 (defined below) also impact the quantum state's stability. Furthermore, each quantum computer has a readout assignment error, and anharmonicity which impacts the quantum state of a qubit \cite{hsia2021physically, Youssef2020}. Physical parameters of various quantum hardware are calibrated and presented in Table \ref{table:param}.
	
	\textit{T2 Time} is the duration of a qubit's quantum state in the excited state. A longer duration of T2 time implies the quantum state stability with minimal loss of information. Higher coherence time indicates better qubit quality for quantum computation.
	
	\textit{T1 Time} is the energy relaxation time of a qubit to lose its quantum state from $\ket{1}$ to $\ket{0}$ when a qubit interacts with the environment. T1 can also be defined as the thermal relaxation time for a qubit to reach the ground state.
	
	\textit{Resonant Frequency} of a qubit can be determined using the Ramsey experiment which determines the transition of a superconducting qubit from $\ket{0}$ to $\ket{1}$ based on the amplitude and duration of a microwave pulse applied to the qubit. The frequency is determined by the energy gap between two quantum states.
	\begin{table*}[htbp]
		\centering	
		\caption{Calibrated Physical Parameters of Quantum  Hardware}
		\label{table:param}	
		\begin{subtable}[h]{1\textwidth}
			\caption{ibm\_osaka Physical Parameters (\textbf{Calibrated at 5:30 PM on 6/19/24})}
			\centering
			\begin{tabular}{|m{2cm}| m{1.5cm}|m{1.7 cm}| m{2.3 cm}|}
				
				\hline
				\textbf{Qubit}&\textbf{T1 (us)}&\textbf{T2(us)}&\textbf{Frequency (GHz)}\\ 
				\hline
				\hline
				Qubit 0 & 403.8&238.66 & 4.718\\
				\hline
				Qubit 1 &320.35 &363.44 & 4.8\\
				\hline
				Qubit 2 &228.67 & 191.89&4.833 \\
				\hline
				Qubit 3 & 350.98&198.82 &4.661 \\
				\hline
				Qubit 4 &43.98 &83.54 & 4.907\\
				\hline
				Qubit 5 & 161.3&68.47 &4.72 \\
				\hline
				Qubit 6 & 334.1& 39.52&4.635 \\
				\hline
				Qubit 7 & 269.28 & 9.07&4.717\\
				\hline
			\end{tabular}	
			\vspace{5 pt}	
		\end{subtable}
		
		\begin{subtable}[h]{1\textwidth}
			\caption{ibm\_kyoto Physical Parameters (\textbf{Calibrated at 1:20 AM on 6/21/24})}
			\centering
			\begin{tabular}{|m{2cm}| m{1.5cm}|m{1.7 cm}| m{2.3 cm}|}
				\hline
				\textbf{Qubit}&\textbf{T1 (us)}&\textbf{T2(us)}&\textbf{Frequency (GHz)}\\ 
				\hline
				\hline
				Qubit 0 & 184.74&30.2 & 4.908\\
				\hline
				Qubit 1 &195.1&71.18 & 4.856\\
				\hline
				Qubit 2 &247.98&51.49&4.733 \\
				\hline
				Qubit 3 & 94.89&47.87 &4.82 \\
				\hline
				Qubit 4 &417.88 &67.25 & 4.854\\
				\hline
				Qubit 5 & 189.22&331.44 &4.728 \\
				\hline
				Qubit 6 & 213.21& 263.22&4.783 \\
				\hline
				Qubit 7 & 329.81 &126.1&4.944\\
				\hline
			\end{tabular}
			\vspace{5 pt}
		\end{subtable}
		\begin{subtable}[h]{1\textwidth}
			\caption{ibm\_sherbrooke Physical Parameters (\textbf{Calibrated at 1:30 AM on 6/21/24})}
			\centering
			\begin{tabular}{|m{2cm}| m{1.5cm}|m{1.7 cm}| m{2.3 cm}|}
				\hline
				\textbf{Qubit}&\textbf{T1 (us)}&\textbf{T2(us)}&\textbf{Frequency (GHz)}\\ 
				\hline
				\hline
				Qubit 0 & 375.14&172.24& 4.636\\
				\hline
				Qubit 1 &351.25&70.13 & 4.736\\
				\hline
				Qubit 2 &237.88&150.08&4.819 \\
				\hline
				Qubit 3 & 370.71&163.46 &4.747 \\
				\hline
				Qubit 4 &120.26 &199.6 & 4.788\\
				\hline
				Qubit 5 & 104.23&161.77&4.851 \\
				\hline
				Qubit 6 & 312.87& 186.59&4.9 \\
				\hline
				Qubit 7 & 120.79 &221.67&4.756\\
				\hline
			\end{tabular}
		\end{subtable}	
		
	\end{table*}

	\section{QPUF 2.0: Proposed Quantum SbD Framework for Smart Grid}
	\label{sec:QPUF-2}
	This section clearly outlines the QSbD approach using QPUF for SCADA-Smart Grid security in detail. It also outlines a comprehensive approach for Smart Grid cybersecurity that can ensure reliability.
	
	The security and reliability of SCADA-enabled critical grid operations and management are essential to counter any hardware/software-based security attacks and vulnerabilities. The data transfer from IEDs at various stages of the grid cycle to an RTU and further to an MTU for analysis needs to be secure to ensure the integrity of data. An adversary may be capable of gaining unauthorized access to an RTU and controlling the on-field IEDs and intelligent protective relays thereby spoofing the IEDs and installing fake nodes. The fake IEDs are installed to provide a false state of a system which could jeopardize the critical grid management and lead to blackouts. Also, software attacks could be performed to access the RTUs controlling IEDs in various subsystems and communication could be eavesdropped due to the unreliable communication framework. The proposed QPUF provides real-time data integrity verification from IEDs by securely attesting each IED in the grid infrastructure through QPUF. The assumptions made in this research paper are:
	\begin{itemize}
		\item All RTUs have access to quantum hardware resources and can obtain QPUF generated responses.
		\item The communication between IED and RTU is
		secure and fool-proof through Transport Layer Security (TLS) communication protocol.
		\item A QPUF evaluated on a noiseless quantum system with enhanced coherence and minimal cross-talk.
		
	\end{itemize}
	All IEDs are resource-constrained, while RTUs, MTUs, and control centers have memory and data processing capabilities. The objective is to develop a secure device attestation scheme using Quantum SbD principles. 
	
	In this work, a smart SCADA-enabled E-CPS security framework is proposed. In the proposed architecture, RTUs, MTUs, and control centers of SCADA are the quantum gateways with efficient access to quantum systems. Various smart electronic devices or IEDs and protective relays can be controlled by RTUs. The RTUs are further controlled by MTUs for data storage, processing, and analysis and work in a master-slave relationship. MTUs act like servers for controlling RTUs at various subsystems in the energy distribution framework. RTUs work as slaves coordinating various field devices and protective relays at geographically distant locations in the energy cycle.
	
	The control center is the command center with an effective human-machine interface and decision-making capability. In the proposed framework, RTUs with reliable access to quantum systems act as gateways for IEDs to access quantum system resources and obtain QPUF generated responses for security. A set of IEDs at a particular location can be securely controlled by an RTU at that specific substation. IEDs can be securely accessed and controlled by all the RTUs with quantum resources in the generation, transmission, and distribution subsystems. MTUs are centralized data processing control systems monitoring RTUs at different substations and energy generation sources for applications like on-field power quality metric evaluation and protective relaying which are critical grid functionalities. Furthermore, MTUs can also authenticate RTUs and perform sensing and actuation processes for grid protection and grid equipment management. MTUs can securely establish communication with RTUs using Quantum Key Distribution (QKD). 
	
\begin{figure}[htbp]	
		\centering
		\includegraphics[width=0.9\textwidth]{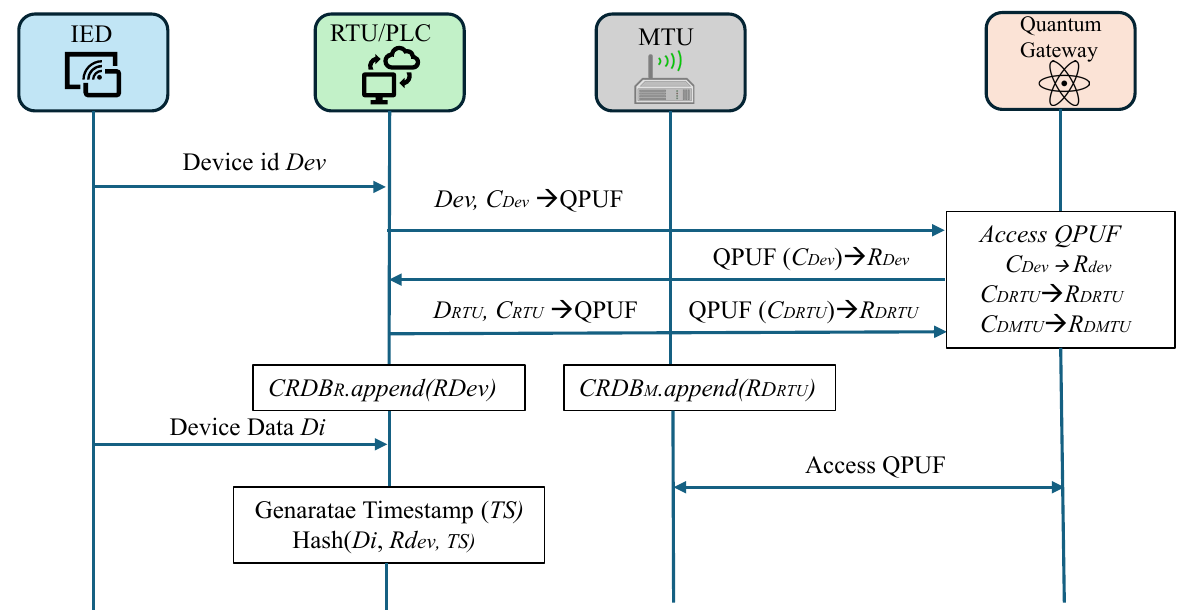}
		\caption{QPUF-based IED Attestation Scheme }
		\label{fig:IED}
	\end{figure}

	In the proposed work, all SCADA entities can access a unique quantum resource or hardware for security and quantum information processing applications, as shown in Fig. \ref{fig:IED}. An IED can send the enrollment request using a unique device identifier device $id$ to an RTU which can be the MAC address. RTUs can be the QPUF gateways with secure access to quantum systems and test the QPUF using challenge initialization states. Each RTU can access a unique quantum system or hardware $QdR$ or $Qx$ for information processing and workflow. Once an RTU receives the enrollment request, a random challenge state $ChallengeCDev$ is selected as input to the QPUF design $Qd$ at the quantum device $Qx$ and $ResponseRDev$ is generated by $Qx$$\rightarrow$$getPUF(CDev)$$\rightarrow$$RDev$. Furthermore, each RTU can communicate with an MTU for coordination and control. An MTU can be secured by obtaining a unique fingerprint $RDMTU$ from device $QdM$$\rightarrow$$getPUF(CDMTU)$$\rightarrow$$RDMTU$. RTUs store the challenge states of a group of IEDs in a secure database $CRDB(CDRTU)$. The MTUs with on-board storage will append the QPUF-generated signatures of all the RTUs in the encrypted form inside a secure database. For storing the signatures, a robust quantum-assisted distributed ledger with an efficient access control mechanism could be adopted. Additionally, each time the communication is established between an IED and RTU after attestation, a unique timestamp $TS $ is generated and the data $di$ from the IED is hashed $ H(di, RDev, TS)$ to ensure integrity. 
	
	\section{Experimental Results}
	\label{sec:ExperimentalResults}
	This section presents the experimental validation details of QPUF 2.0 along with its performance evaluation and analysis in detail.
	\subsection{Experimental Setup}
	Experimental evaluation was performed on quantum systems and simulators from IBM and Google. IBM quantum systems are accessible to users through an application programming interface token that is loaded each time a user accesses the quantum system. A user can access a limited number of quantum systems with an open plan and an available qiskit run time of 10 minutes. All the IBM quantum simulators are available for free access to test and execute quantum algorithms with ample run time. 
	
\begin{figure}[htbp]	
		\centering
		\includegraphics[width=0.9\textwidth]{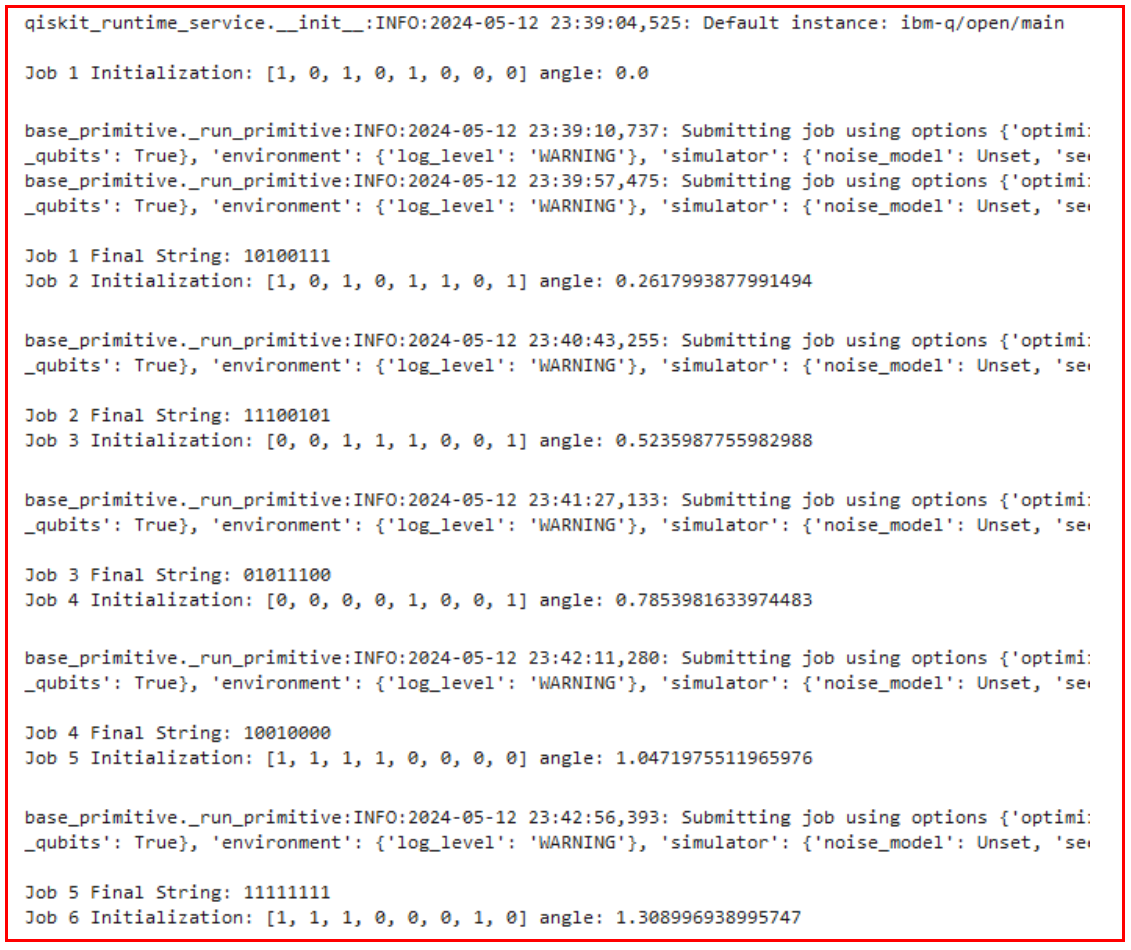}
		\caption{Executing QPUF circuit on IBM quantum system and extracting QPUF responses}
		\label{fig:Qubit}
	\end{figure}

	For the evaluation, QPUF was deployed both on the simulator and quantum hardware. IBM's "ibmq\_qasm\_Simulator" is chosen for QPUF design evaluation on a simulator. For hardware performance evaluation, IBM's "ibm\_osaka" 'ibm\_kyoto' and "ibm\_sherbrooke" quantum hardware with Eagle R3 processor supporting 127-qubits are chosen. Eagle R3 is an advanced quantum processor with enhanced coherence time facilitating complex computations. The proposed QPUF design is programmed using Python and is tested on IBM's resources using Qiskit, a quantum computing platform to evaluate and execute circuits and algorithms on IBM quantum systems \cite{qiskit2024}. Our proposed design is an 8-qubit architecture with 8 quantum and 8 classical bits tested with Hadamard, CNOT, Ry, and Idle gates. Overall, 75 quantum jobs are executed with 1024 shots for each job on the IBM quantum simulator. QPUF circuit is tested with different tunable rotation Ry gate angles and qubit initialization states as challenge input. For each job, a rotation angle in the range of 0-2 pi that places each qubit in an unknown quantum state is tested as challenge input to QPUF. All the qubits in the QPUF are tested with a unique initialization state either 0 or 1 for each qubit and a tunable rotation Ry gate angle is chosen and applied to all the qubits in the circuit for each execution or job. The QPUF then generates a unique final string as a response for each execution as presented in Fig. \ref{fig:Qubit}.
	
\begin{figure}[htbp]
		\centering
		\begin{subfigure}{0.5\textwidth}
			\centering
			\includegraphics[width=\textwidth]{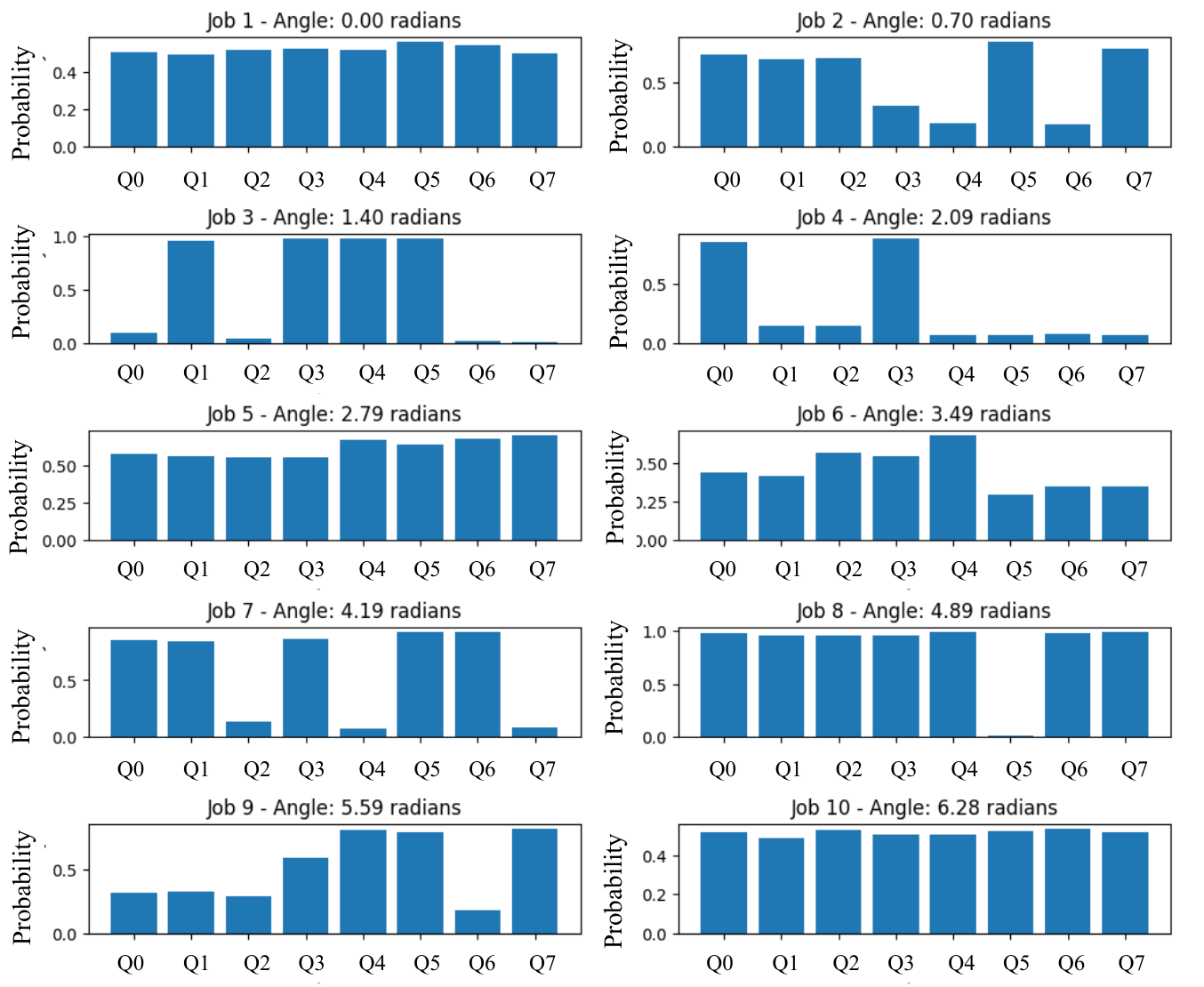}
			\caption{ibm\_osaka}		
		\end{subfigure}
		
		\begin{subfigure}{0.495\textwidth}
			\centering
			\includegraphics[width=\textwidth]{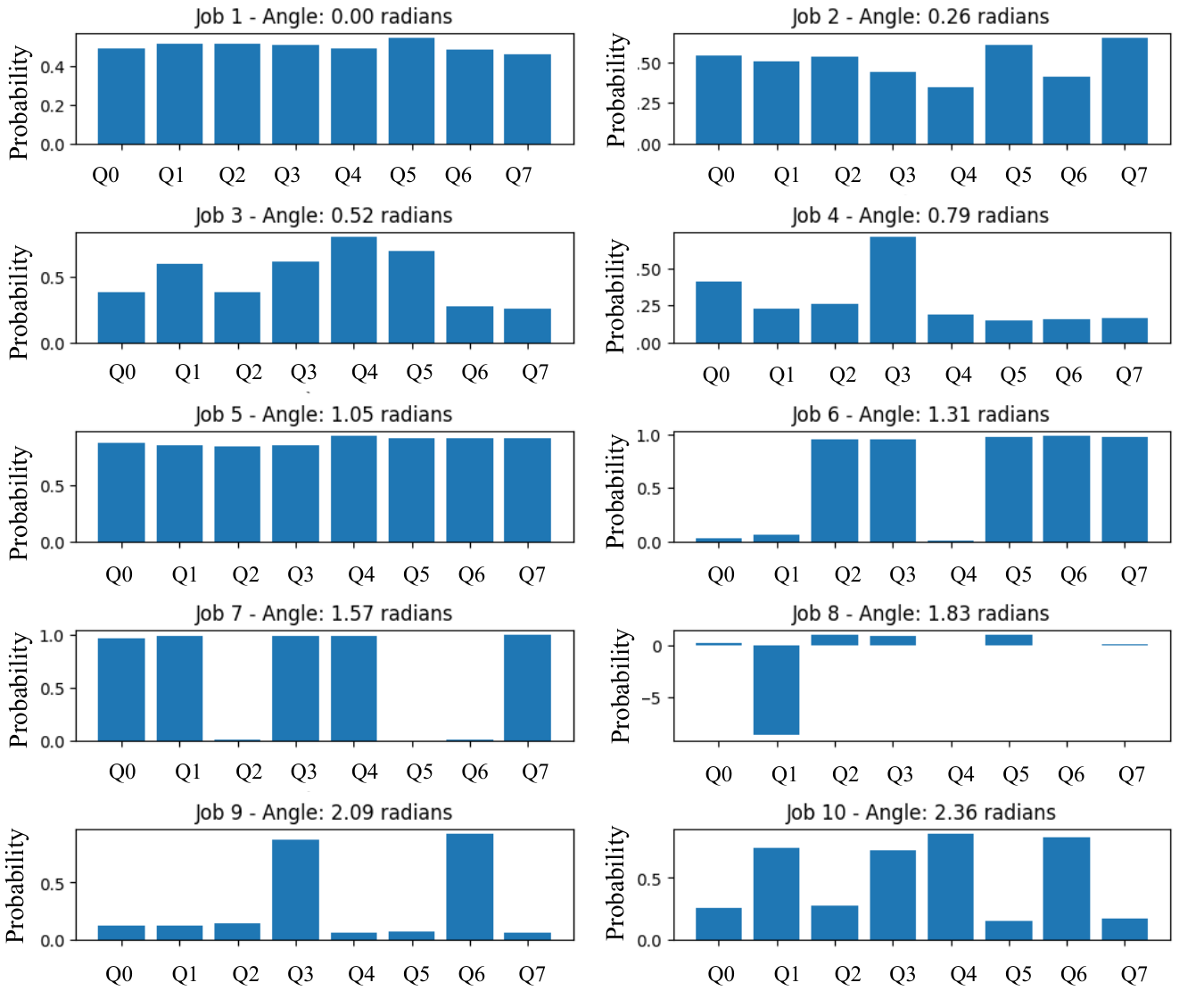}
			\caption{ibm\_kyoto}
		\end{subfigure}
		\begin{subfigure}{0.495\textwidth}
			\centering
			\includegraphics[width=\textwidth]{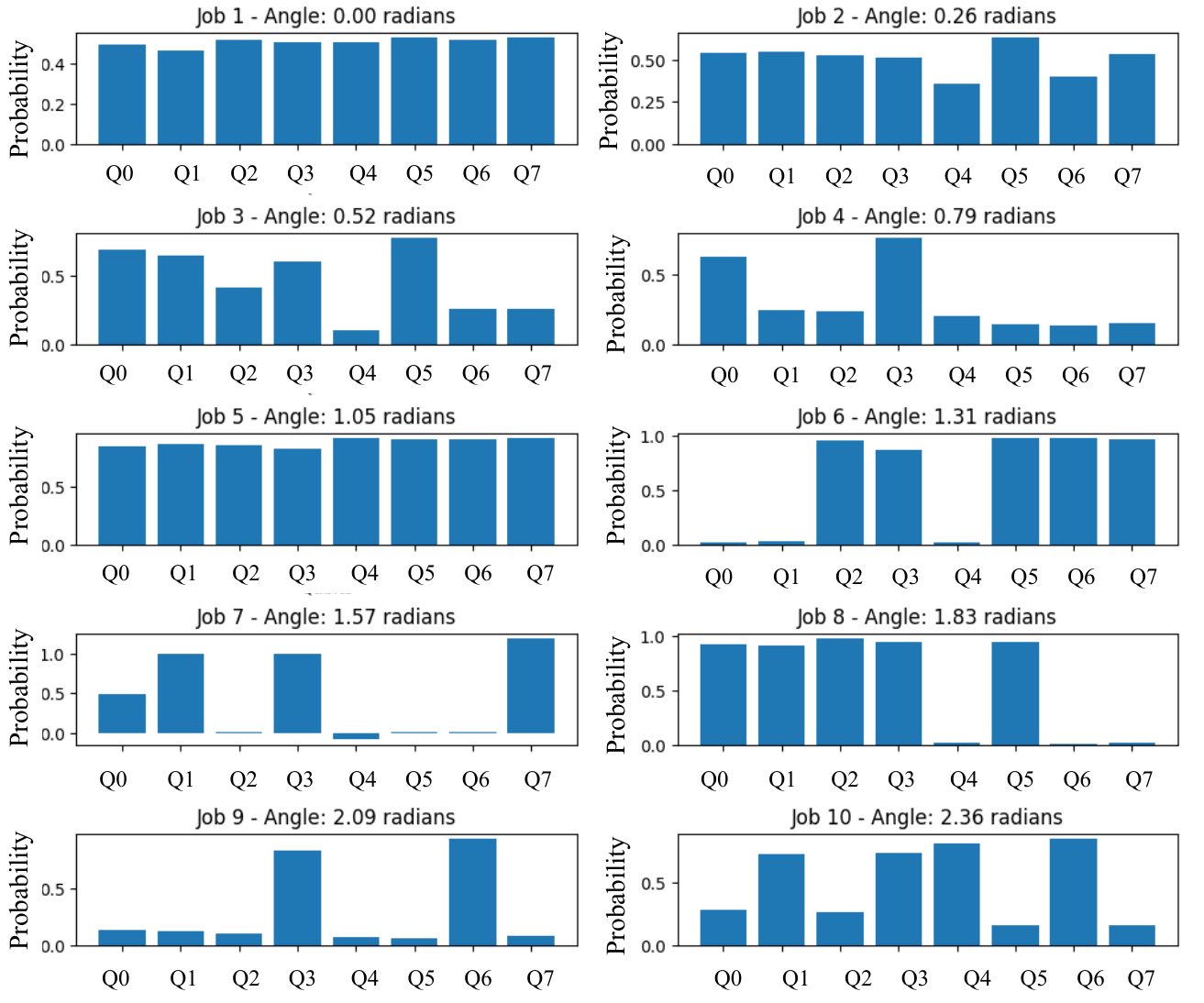}
			\caption{ibm\_sherbrooke}
		\end{subfigure}	
		
		\caption{Qubit's state probabilities for the QPUF Circuit from Various Quantum Systems}
		\label{fig:osaka}		
	\end{figure}

\begin{figure*}[htbp]	
		\centering
		\includegraphics[width=0.75\textwidth]{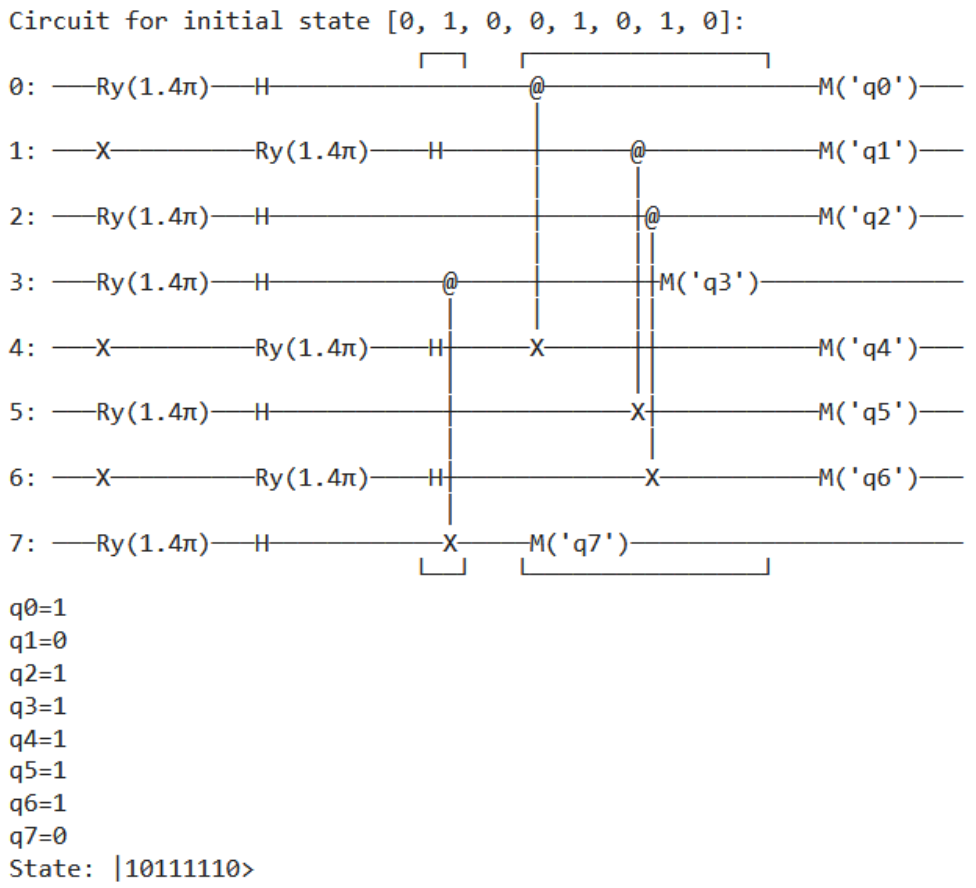}
		\caption{QPUF Evaluation on Google Quantum Simulator (cirq.sim.sparse\_simulator)}
		\label{fig:cirq-output}	
	\end{figure*}

	The QPUF design is evaluated for five instances on a simulator with 75 executions for each instance to evaluate reliability.  Also, the QPUF circuit is evaluated on ibm\_osaka, ibm\_kyoto, and ibm\_sherbrooke hardware for 3 instances with 10 circuit executions with 1024 shots in each instance. The open plan of IBM supports a qiskit run time of 10 minutes which allows only a limited number of circuit executions and shots. A shot in the quantum circuit execution refers to the number of measurement outcomes. Each circuit execution or job obtains 1024 measurement outcomes or shots for the circuit. Therefore, a QPUF job execution implies testing the QPUF with a challenge state and tunable rotation angle for 1024 executions. The qubit's 1 state probability for each job in the QPUF circuit execution with 1024 shots on the quantum hardware is presented in Fig. \ref{fig:osaka}. The QPUF is also evaluated on Google quantum simulator using 'Cirq'. Cirq is a Python framework for quantum algorithm and circuit evaluation on Google quantum systems and simulators \cite{CirqDevelopers2024}. The proposed design was evaluated on Cirq's 'cirq.simulator()' which is considered effective for flexibility and can emulate quantum hardware behavior. The circuit was evaluated for 100 jobs in each instance with varying tunable rotational angles unique for each circuit or job execution. The circuit was evaluated on the Google simulator and metrics are evaluated and presented in Fig. \ref{fig:cirq-output} for comparative performance analysis with the IBM quantum simulator.
	
	\begin{figure*}[ht]
		\centering
		\setlength{\tabcolsep}{0pt}
		\begin{subfigure}{1\linewidth}
			\includegraphics[width=1\textwidth]{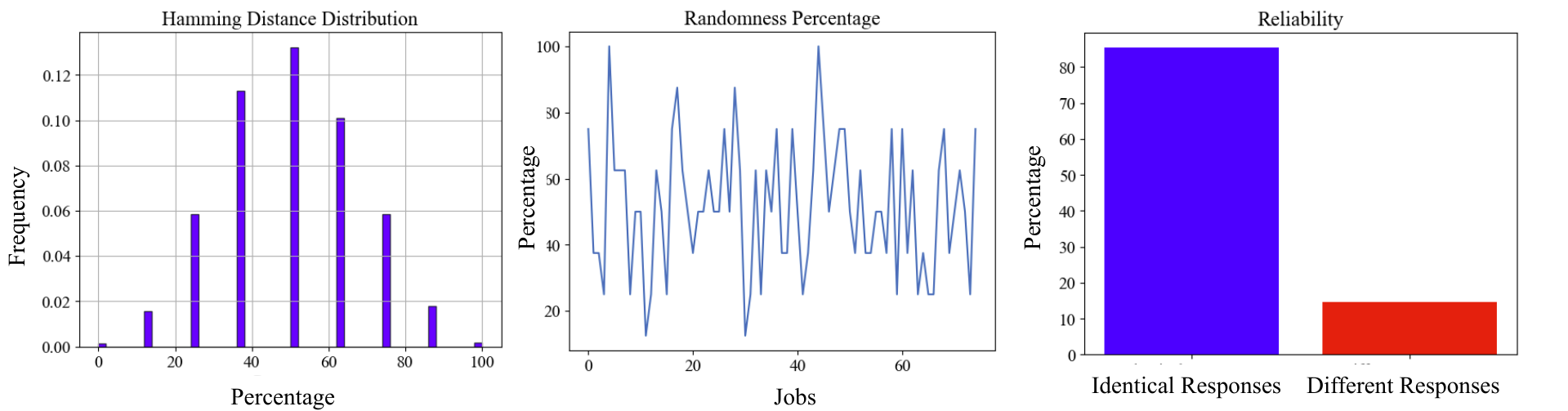}
			\caption{QPUF Evaluation Results from IBM Simulator (ibmq\_qasm\_simulator)}
		\end{subfigure} 
		\begin{subfigure}{1\linewidth}
			\includegraphics[width=1\textwidth]{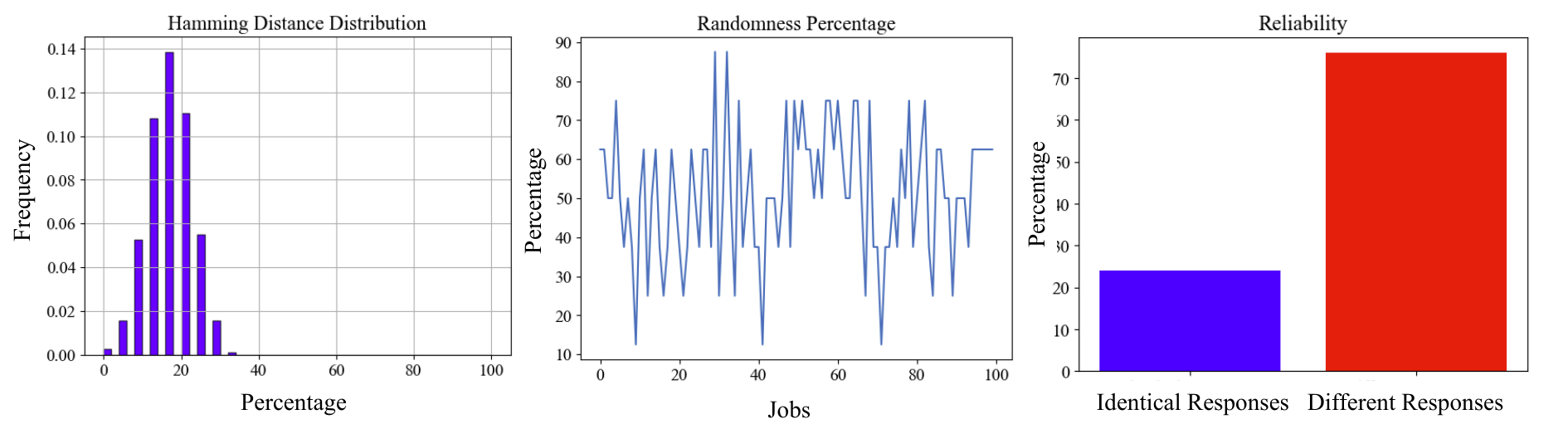}
			\caption{ QPUF Evaluation Results from Google Quantum Simulator}
		\end{subfigure}
		\caption{QPUF Performance Evaluation Metrics}
		\label{fig:qsimulator}		
	\end{figure*}
	
	The QPUF metrics evaluated on the obtained job execution outcomes are defined and presented below:
	
	\textit{Randomness} of a PUF is the measure of balance in the occurrence of ones and zeros in a given response. For a given QPUF response $ri$ the randomness is obtained by counting the number of one's $ki$ in $ri$ and dividing it by the total number of bits $bi$ in $ri$. 
	
	\begin{center}
		Randomness$(ri\%)= \frac{ki}{bi}\times 100$
	\end{center}
	
	\textit{Diffuseness} of a QPUF is the extent of variation of a QPUF response with varying initialization states and angles. In the context of QPUF, Diffuseness is obtained by calculating the average intra-hamming distance of QPUF responses from a hardware or simulator with varying challenge states. To evaluate the diffuseness of QPUF on a device, intra-hamming distance is calculated among responses $r1$ and $r2$ for tunable angles $a1$, $a2$ and quantum state initializations $i1$, $i2$.
	\begin{center}
		Diffuseness ($r1$,$r2$) = $HD$(r1, r2);
	\end{center}  
	
	The reliability of a QPUF is the extent of matching of QPUF responses under noisy and environmental conditions impacting the quantum systems. To evaluate reliability, Hamming distance HD ($j1$,$j2$ ) of two instances of QPUF executions from a quantum system is calculated. The final response (rn) from both instances achieving 100\% reliability with no varying bits is considered reliable. 
	\begin{center}
		Reliability(\%) = 100 - HD(j1, j2)
		\begin{equation}
			rn = \begin{cases}
				100\%, & \text{if } HD = 0, \\
				0\%, & \text{if } HD \neq 0.
			\end{cases}
		\end{equation}
	\end{center}
	
	The uniqueness of a QPUF is the degree of variation of QPUF responses on various hardware. To calculate QPUF uniqueness, the average hamming distance of execution instances $rn$, and $r_{n+1}$ from various hardware $s_{n}$ and $s_{n+1}$ are calculated and can be represented as:
	
	\begin{center}
		Uniqueness ($s_{n}$, $s_{n+1}$) =  HD ( $rn$, $r_{n+1}$)x 100
	\end{center}
	The QPUF metrics evaluated for IBM and Google quantum systems and simulators are presented in Fig. \ref{fig:qsimulator}. 
	\subsection{Pulse level Control of QPUF}
	The quantum circuit logic at the higher level translates to microwave pulse level control of superconducting quantum circuits at hardware. The microwave pulse at a specified amplitude, phase, and duration manipulates the quantum state of a qubit. The desired performance of a circuit implementation can be realized through careful calibration of microwave pulses controlling the quantum hardware. IBM’s qiskit Pulse is an open-source front-end implementation interface for IBM quantum systems. Channels in pulse level control drive microwave pulses to manipulate the quantum state of physical qubits in superconducting circuits \cite{Alexander2020}. Qiskit pulse includes various channels that facilitate qubit interaction for diverse functions which include measurement, control, state change, and state acquisition. Various channels in the qiskit pulse include:
	
	\textit{Drive Channels(Di):} This channel is connected to a qubit where a microwave pulse in resonance with the qubit's frequency can be applied to control the qubit's quantum state.
	
	\textit{Measurement Channel (Mi):} This channel enables microwave pulse level interaction with readout detectors to perform qubit’s state measurement after executing the quantum circuit.
	
	\textit{Acquire Channel(Ai):} This channel enables efficient readout of quantum state by digitizing and converting the measurement data into a suitable format.
	
	\textit{Control channel(Ui):} Control channels enable efficient implementation of quantum circuits by sending control signals to various components of quantum hardware. 
	
\begin{figure*}[htbp]
		\centering
		\begin{subfigure}{1\textwidth}
			\centering
			\includegraphics[width=\linewidth]{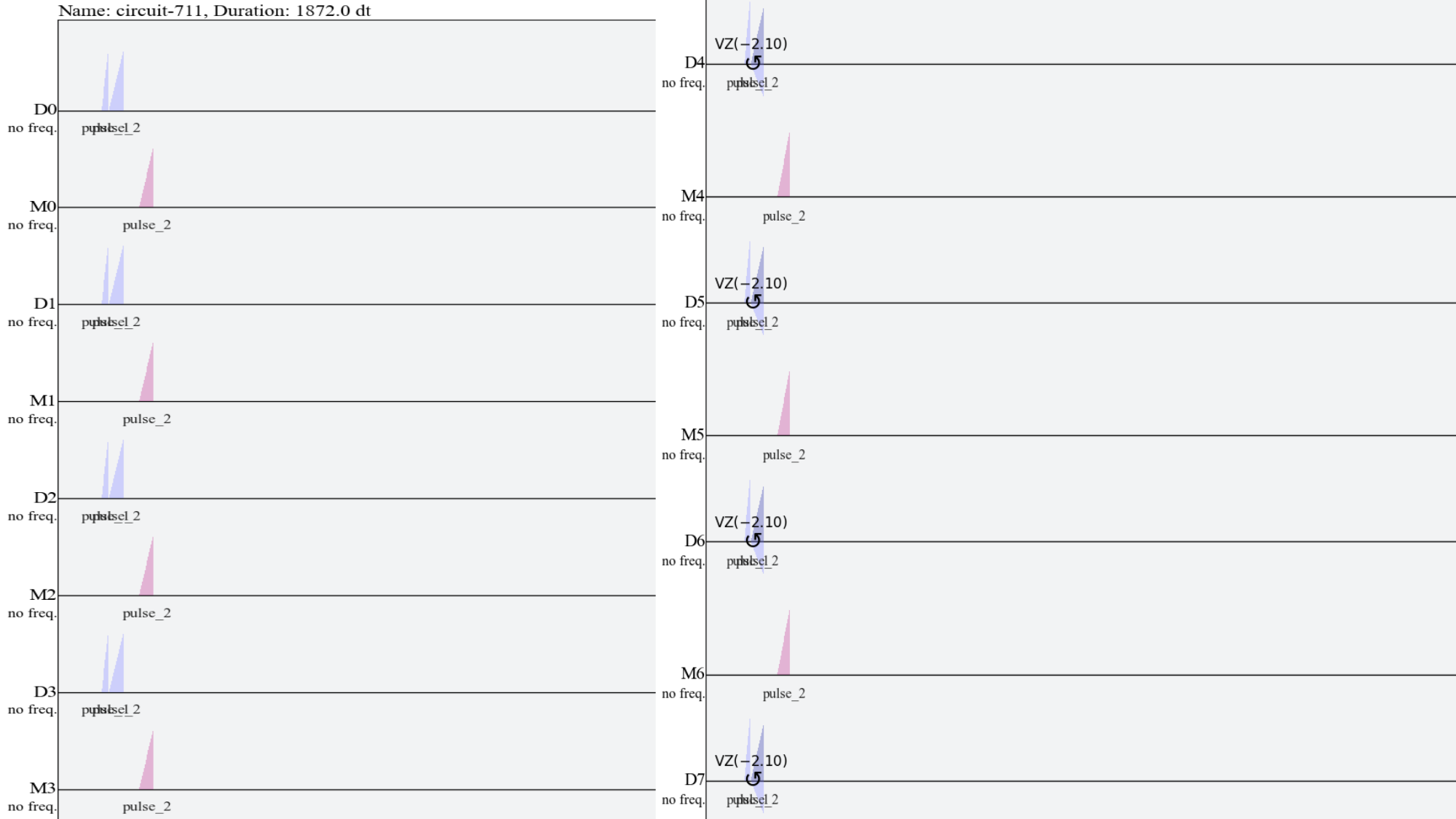}
			\caption{Qiskit Pulse Drive and Measurement Channels for 8 Qubits}
		\end{subfigure}
		
		\begin{subfigure}{0.75\textwidth}
			\centering
			\includegraphics[width=\linewidth]{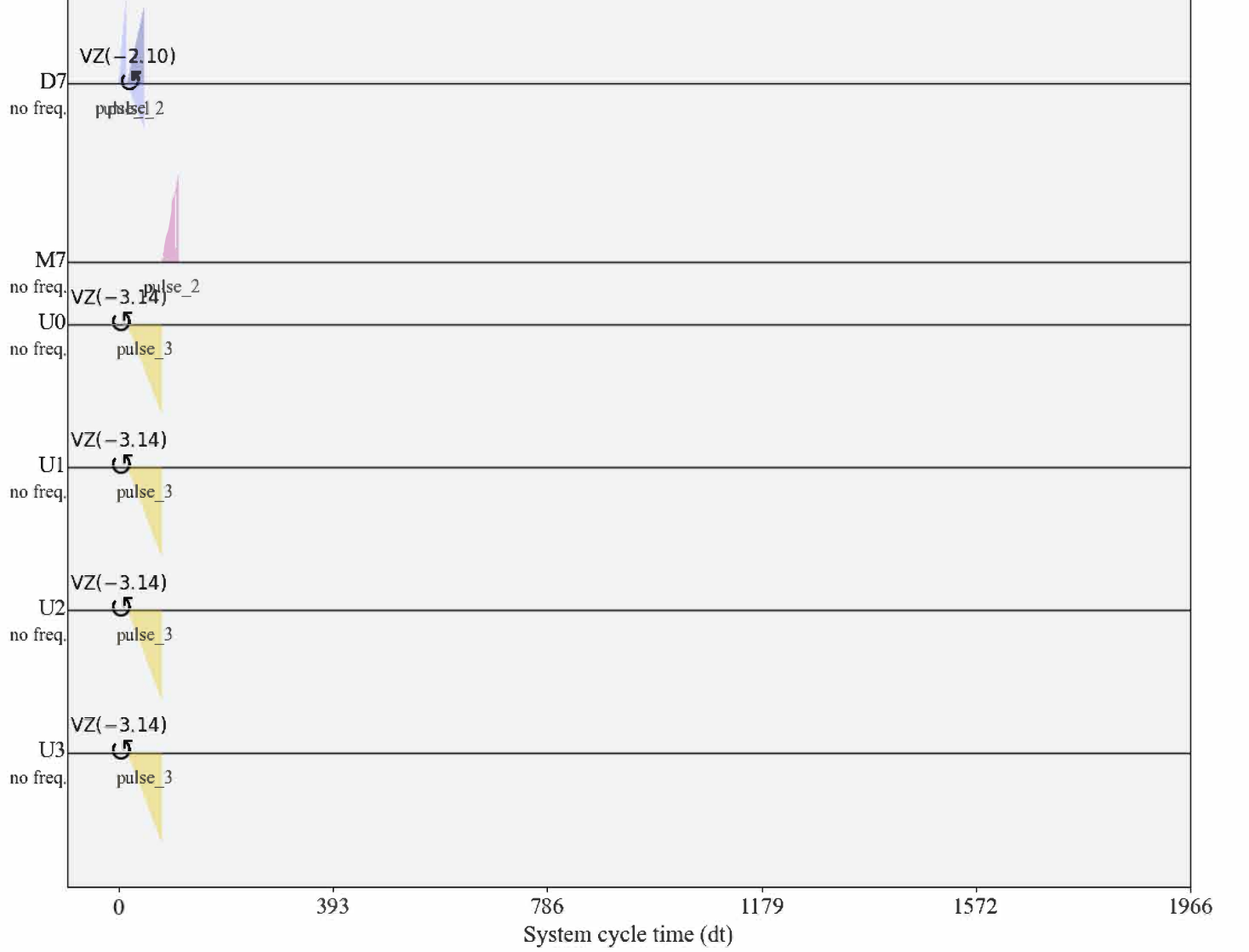}
			\caption{Qiskit Pulse Control Channels for Entangled Qubits}
		\end{subfigure}

		\caption{Qiskit Pulse Schedule for QPUF}
		\label{fig:pulse}		
	\end{figure*}

	Furthermore, Specific pulse level instructions are used to calibrate the pulse amplitude, phase, control, and delay to perform quantum state changes which include Delay, Play, ShiftPhase, ShiftFrequency, and Acquire.
	\begin{itemize}
		\item Delay instruction adds or idles the channel for a specified time while Play instruction gives the pulse output waveform of the channel.  
		\item ShiftPhase and ShiftFrequency shifts the phase and frequency of pulse on the channel to effect a change in the qubit's state.
		\item Acquire instruction collects the measurement results and stores them in a classical register.
	\end{itemize}
	For pulse level demonstration of QPUF circuit, '\textbf{GenericBackendV2}', a customized backend with 8 qubits from qiskit-aer is chosen. It mimics the behavior of real quantum hardware while providing a pulse-level control behavior for the circuit. Fig. \ref{fig:pulse} shows the pulse schedule of the QPUF with 8 qubits. The shape and duration of each pulse represent the quantum gate operations on the qubits. The pulse schedule has 8 drive and measurement channels for the 8 qubits in the quantum circuit. The vz pulse is a virtual microwave pulse at -3.14 radians and is used to remove phase differences in the quantum state ensuring reliable state manipulation.

	\subsection{Analysis and Challenges}
	
	The QPUF circuit execution on hardware could support only fewer job executions for each instance with 1024 measurement samples. This is a very low sample space for performance analysis. The obtained uniqueness of QPUF for various simulators and hardware is around 17\% and performance evaluation as presented in Fig. \ref{fig:qsimulator} and Table \ref{Table:QPUF-Freq} clearly show the potential for further improvement in reliability on ibm\_osaka and other hardware with increasing circuit execution space. The evaluation, however, proves the efficiency of QPUF on the IBM quantum simulator achieving better reliability, randomness, and uniqueness. 
	
	The improved qubit coherence can increase the resiliency of qubits to noise and environment thereby achieving better qubit coherence which can improve the reliability of QPUF. The realization of QPUF and reliable response extraction from noisy quantum computers is a great challenge due to the very low-paced improvement in the realization of noise-free quantum hardware. This research work explored the scope of quantum mechanics principles for QSbD and presented a QPUF architecture that has shown PUF execution on Quantum hardware as a feasible approach with enhanced performance in comparison to the related research. However, to further evaluate the robustness of the proposed QPUF, more experimental space is needed along with increased access to noiseless quantum resources from various companies.
	\begin{table}[htbp]
		\caption{QPUF Performance Evaluation}
		\label{Table:QPUF-Freq}
		\centering
		\begin{tabular}{|M{2.5cm}| M{2.7cm}|M{2.4cm}| M{2.4 cm}|}
			\hline
			\textbf{System}&\textbf{Intra-Hamming Distance}(\%) &\textbf{Randomness}(\%) & \textbf{Reliability(\%)}\\ 
			\hline
			\hline
			{ibmq\_qasm\_ simulator} & 50& 52 &87\\
			\hline
			ibm\_osaka & 48 & 63&40\\
			\hline
			ibm\_kyoto &49 &51&70 \\
			\hline
			ibm\_sherbrooke &47 &56 &60\\
			\hline
			Google Simulator & ~16 & ~51 &~17-24\\
			\hline
		\end{tabular}
	\end{table}
	\section{Conclusion and Future Research}
	\label{sec:conclusion}
	
	This research work validated and presented a novel QSbD framework for reliable authentication of power transmission systems ensuring the security of protective relays and intelligent electronic devices in the SCADA-Smart grid infrastructure. The performance evaluation demonstrates the scope of QPUF for SbD through hardware evaluation. QPUF circuit execution demonstrates the potential for QSbD application in smart grid cybersecurity with a robust authentication framework. The QPUF can be a feasible security solution for smart grid due to its capability to support high-performance computing applications with SCADA enabled energy transmission and distribution infrastructure. The QPUF demonstration on quantum hardware "ibm\_osaka", "ibm\_kyoto", and "ibm\_sherbrooke" have shown a diffuseness, and randomness at around 50\% and achieves a 90\% reliability on the simulator. Furthermore, our experimental validation has shown a PUF challenge response generation mechanism from QPUF design harnessing quantum entanglement, superposition, and decoherence principles along with QPUF design metric evaluation further evaluating its robustness. The proposed QSbD SCADA framework for smart grid can enhance the trustworthiness of various subsystems and their communication in the energy infrastructure.
	
	Furthermore, our analysis shows that the QSbD can be resource intensive for Healthcare-CPS (H-CPS) which requires resource efficient and technically feasible security solutions. As the extent of application of quantum computers is projected to increase in the coming years, the reliability and application of these systems increase improving the information processing capability manifold. Also, the QPUF can be integrated with quantum cryptography protocols based on the principles of QSbD and SbD for enhanced privacy, security, and authenticity for the emerging Quantum IoT era.
	
	\bibliographystyle{unsrt}
	\bibliography{Bibliography_QPUF_2-0}

	\section* {Authors:}
	
	\begin{minipage}{0.2\textwidth}
		\centering
		\includegraphics[width=1.0in, height=1in, clip, keepaspectratio]{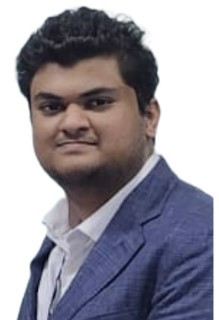}
	\end{minipage}%
	\begin{minipage}{0.75\textwidth}
		\textbf{Vishnu Bathalapalli} is a Doctoral Student in the Smart Electronic Systems Laboratory (SESL) in the Department of Computer Science and Engineering (CSE) under the guidance of Dr. Saraju Mohanty at the University of North Texas (UNT). He earned his Bachelor of Technology (B.Tech) degree in Electronics and Communication Engineering (ECE) from Sri Venkateswara University in Tirupati, India. His research focuses on addressing data and device security challenges through Security-by-Design utilizing Physical Unclonable Functions (PUFs), Trusted Platform Modules (TPMs), Quantum PUFs, and Blockchain. His research aims to incorporate robust security measures early in the development process for applications in Smart Healthcare, Deepfake Mitigation, and Smart Energy. He is an author of over 10 journal and conference publications.
	\end{minipage}
	
	\noindent
	\begin{minipage}{0.2\textwidth}
		\centering
		\includegraphics[width=1.0in, height=1.0in, clip, keepaspectratio]{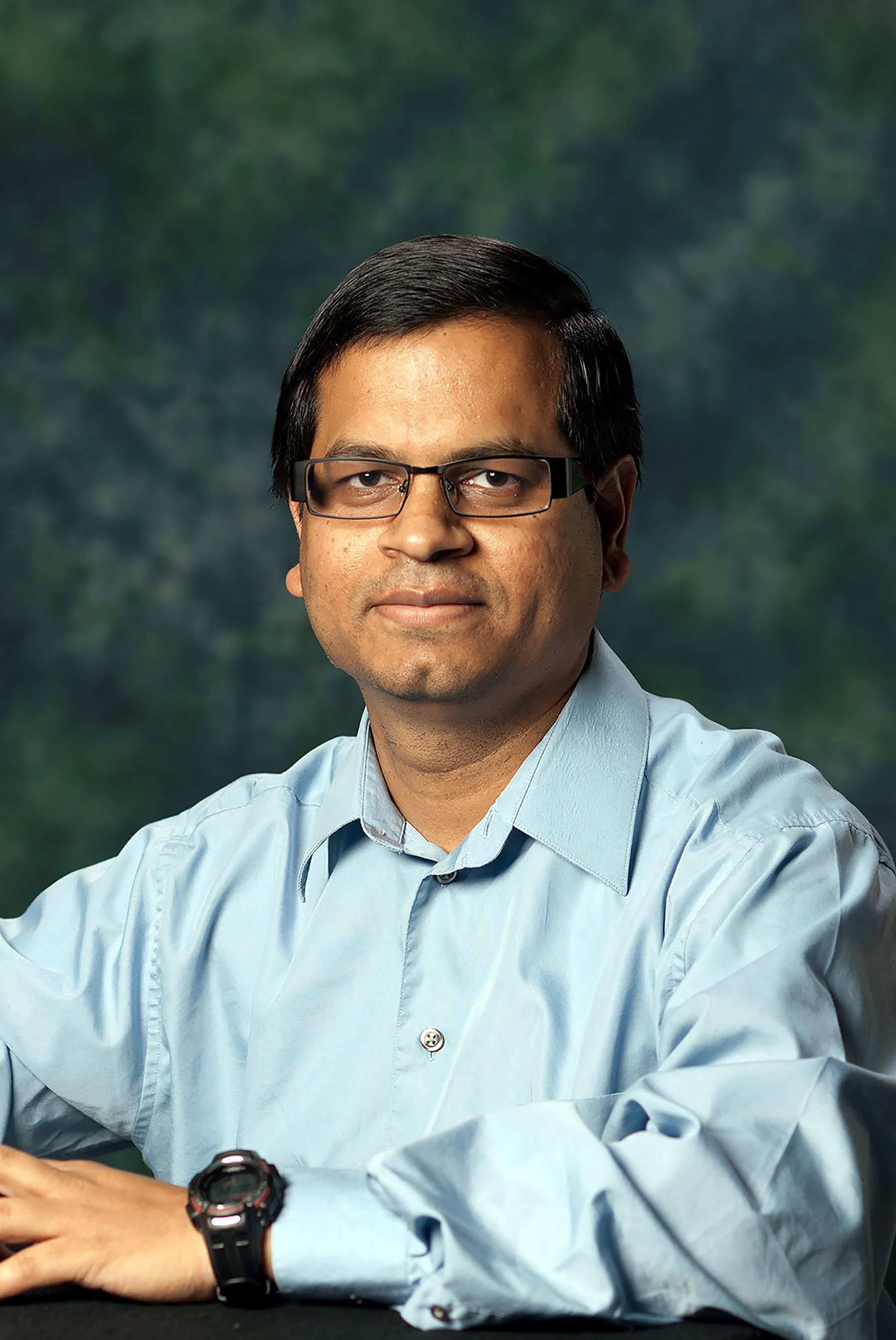}
	\end{minipage}%
	\begin{minipage}{0.75\textwidth}
		\textbf{Saraju P. Mohanty} (Senior Member, IEEE) received the bachelor’s degree (Honors) in electrical engineering from the Orissa University of Agriculture and Technology, Bhubaneswar, in 1995, the master’s degree in Systems Science and Automation from the Indian Institute of Science, Bengaluru, in 1999, and the Ph.D. degree in Computer Science and Engineering from the University of South Florida, Tampa, in 2003. He is a Professor with the University of North Texas. His research is in “Smart Electronic Systems” which has been funded by National Science Foundations (NSF), Semiconductor Research Corporation (SRC), U.S. Air Force, IUSSTF, and Mission Innovation. He has authored 550 research articles, 5 books, and 10 granted and pending patents. His Google Scholar h-index is 58 and i10-index is 269 with 15,000 citations. He is regarded as a visionary researcher on Smart Cities technology in which his research deals with security and energy aware, and AI/ML-integrated smart components. He introduced the Secure Digital Camera (SDC) in 2004 with built-in security features designed using Hardware Assisted Security (HAS) or Security by Design (SbD) principle. He is widely credited as the designer for the first digital watermarking chip in 2004 and first the low-power digital watermarking chip in 2006. He is a recipient of 19 best paper awards, Fulbright Specialist Award in 2021, IEEE Consumer Electronics Society Outstanding Service Award in 2020, the IEEE-CS-TCVLSI Distinguished Leadership Award in 2018, and the PROSE Award for Best Textbook in Physical Sciences and Mathematics category in 2016. He has delivered 30 keynotes
		and served on 15 panels at various International Conferences. He has been serving on the editorial board of several peer-reviewed international transactions/journals, including IEEE Transactions on Big Data (TBD), IEEE Transactions on Computer-Aided Design of Integrated Circuits and Systems (TCAD), IEEE Transactions on Consumer Electronics (TCE), and ACM Journal on Emerging Technologies in Computing Systems (JETC). He has been the Editor-in-Chief
		(EiC) of the IEEE Consumer Electronics Magazine (MCE) during 2016-2021. He served as the Chair of Technical Committee on Very Large Scale Integration (TCVLSI), IEEE Computer Society (IEEE-CS) during 2014-2018 and on the Board of Governors of the IEEE Consumer Electronics Society during 2019-2021. He serves on the steering, organizing, and program committees of several international conferences. He is the steering committee chair/vice-chair
		for the IEEE International Symposium on Smart Electronic Systems (IEEE-iSES), the IEEE-CS Symposium on VLSI (ISVLSI), and the OITS International Conference on Information Technology (OCIT). He has supervised 3 post-doctoral researchers, 17 Ph.D. dissertations, 28 M.S. theses, and 28 undergraduate projects
	\end{minipage}

	\noindent
	\begin{minipage}{0.2\textwidth}
		\centering
		\includegraphics[width=1in, height=1in, keepaspectratio]{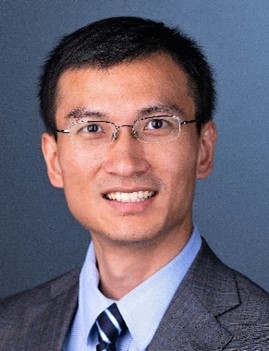}
	\end{minipage}%
	\begin{minipage}{0.75\textwidth}
		\textbf{Chenyun Pan} (Senior Member, IEEE) received a B.S. degree in microelectronics from Shanghai Jiao Tong University, Shanghai, China, in 2010 and a Ph.D. in ECE from Georgia Institute of Technology in 2015. He is currently an Assistant Professor at the Department of Electrical Engineering, The University of Texas at Arlington. His research interests include device-, circuit-, and system-level modeling and optimization for energy-efficient Boolean and non-Boolean computing systems based on various emerging device and interconnect technologies. He has published over 70 peer-reviewed IEEE journal and conference papers. He is the recipient of two Best Paper awards in the IEEE International Symposium on Quality Electronic Design and IEEE Conference on IC Design and Technology, Research Spotlight Award in the School of ECE at Georgia Tech, and early career research program award from US Department of Energy. 
	\end{minipage}

	\noindent
	\begin{minipage}{0.2\textwidth}
		\centering
		\includegraphics[width=1in, height=1in, keepaspectratio]{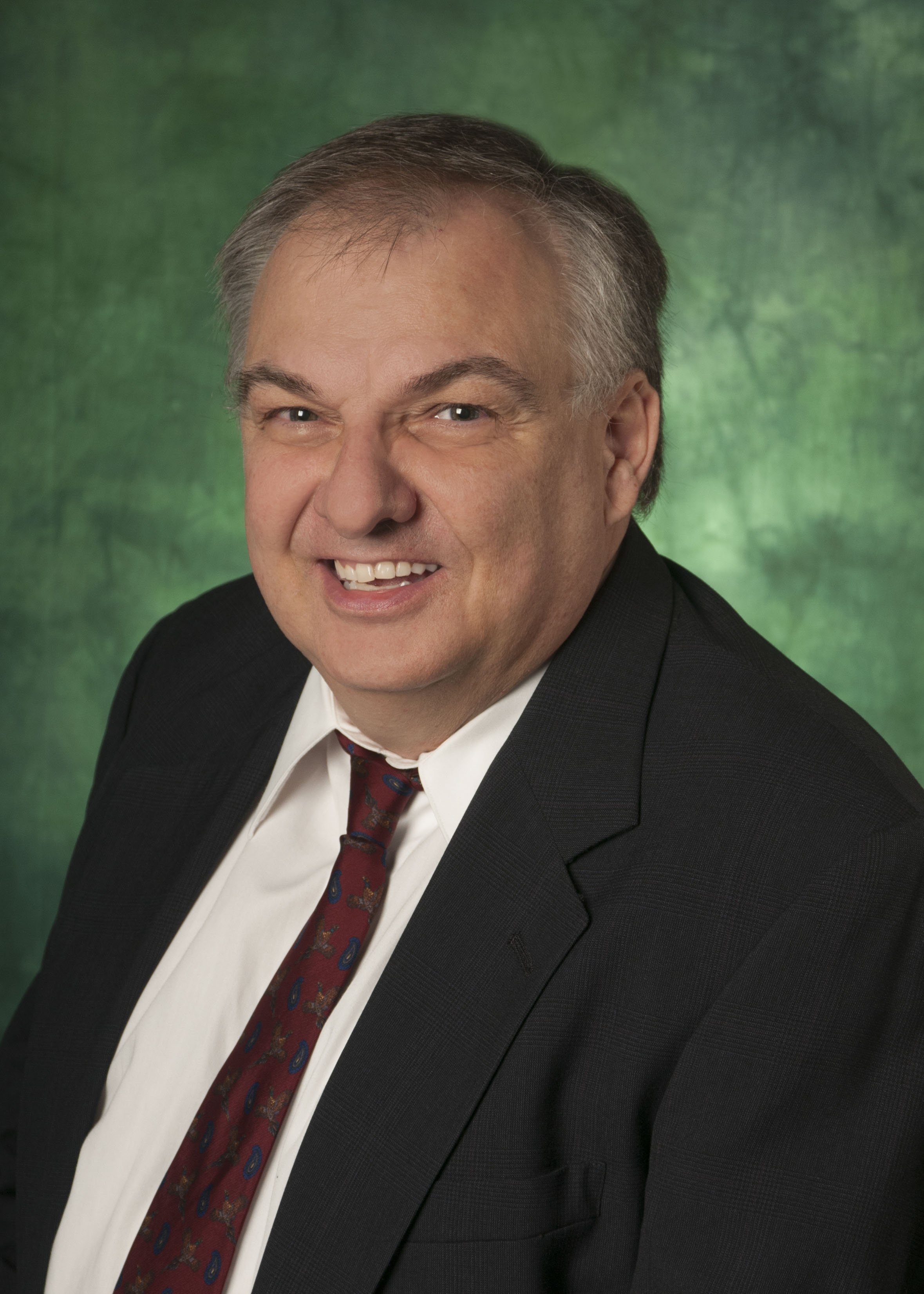}
	\end{minipage}%
	\begin{minipage}{0.75\textwidth}
		\textbf{Elias Kougianos} received a BSEE from the University of Patras, Greece in 1985 and an MSEE in 1987, an MS in Physics in 1988 and a Ph.D. in EE in 1997, all from Louisiana State University. From 1988 through 1998 he was with Texas Instruments, Inc., in Houston and Dallas, TX. In 1998 he joined Avant! Corp. (now Synopsys) in Phoenix, AZ as a Senior Applications engineer and in 2000 he joined Cadence Design Systems, Inc., in Dallas, TX as a Senior Architect in Analog/Mixed-Signal Custom IC design. He has been at UNT since 2004. He is a Professor in the Department of Electrical Engineering, at the University of North Texas (UNT), Denton, TX. His research interests are in the area of Analog/Mixed-Signal/RF IC design and simulation and in the development of VLSI architectures for multimedia applications. He is an author of over 200 peer-reviewed journal and conference publications.
	\end{minipage}
	
\end{document}